\documentstyle[aps,epsfig]{revtex}

\graphicspath{{figures/}}
\begin{document}
\draft      
%
%

\newcommand{\txt}[1]{{\mathrm{#1}}} 
\newcommand{\ppORppbar}{pp\!\!\!\!^{^{(-)}}}
\newcommand{\qORqbar}{q\!\!\!\!^{^{(-)}}}
\newcommand{\order}{\mbox{${\cal O}$}}
\newcommand{\Matrix}{ {\cal{M}} }
\newcommand{\const}{\mbox{\sc\small Const}}
\newcommand{\nbody}{n-body}
\newcommand{\nplusbody}{(n+1)-body}
\newcommand{\smin}{\mathrm{s}_{\mathrm{min}}}
\newcommand{\szero}{\mathrm{s}_{\mathrm{zero}}}
\newcommand{\sPS}{\mathrm{s}_{\mathrm{P.S.}}}
\newcommand{\NLOa}{NLO}
\newcommand{\deltaS}{\delta_\txt{s}}
\newcommand{\deltaC}{\delta_\txt{c}}
\newcommand{\PSS}{PSS}
\newcommand{\PSveto}{$\Phi$-space~Veto}
\newcommand{\MSbar}{\overline{\mathrm{MS}}}
\newcommand{\HERWIG}{{\tt HERWIG}}
\newcommand{\PYTHIA}{{\tt PYTHIA}}
\newcommand{\ISAJET}{{\tt ISAJET}}
\newcommand{\CXX}{C++}
\newcommand{\mass}[1]{M_\txt{#1}}
\newcommand{\pT}[1]{P^T_\txt{#1}}
\newcommand{\alphaS}{\mbox{$\alpha_\txt{S}$}}
\newcommand{\alphaQED}{\mbox{$\alpha_\txt{QED}$}}

\newcommand{\XSECTESTRESULT}{$239.7\pm0.6$~pb}

\topmargin 0.in


%
%
\title{Phase space veto method for next-to-leading order event
 generators in hadronic collisions}

\author{Matt Dobbs\footnote{Electronic Mail Address: matt.dobbs@cern.ch} }
\address{
  Department of Physics and Astronomy, University of Victoria,
  P.O.\ Box 3055, Victoria, British Columbia, Canada V8W~3P6}
\date{November 10, 2001}
\maketitle

\begin{abstract}

A method for organizing next-to-leading order QCD calculations using
a veto which enforces the cancellations between virtual and real
emission diagrams is applied to hadronic collisions.
The method employs phase space slicing with the slicing parameter
determined dynamically event-by-event. It allows for the generation of
unweighted events and can be consistently merged with a parton shower.
The end product is more intuitive for the end user, as it is
probabilistic, and can be easily interfaced to general purpose
showering and hadronization programs to obtain a complete event
description suitable for experimental analyses.  As an example an
event generator for the process $\ppORppbar\rightarrow Z + X$ at
\NLOa\ is presented and interfaced consistently to the \PYTHIA\ shower
and hadronization package.

\end{abstract}

%
%


%
%
\section{ INTRODUCTION } \label{s_introduction}

Computer simulations of higher order quantum chromodynamics (QCD)
corrections in high energy physics collisions typically rely on one of
two methods: the parton
shower~\cite{Sjostrand:1985xi,Marchesini:1988cf} or numerical
integration of next-to-leading order (\NLOa) distributions. Both
methods have proven extremely useful, but each has a limited---though
complementary---region of applicability.

Programs which employ the parton shower approach, such as
\PYTHIA~\cite{Sjostrand:2001wi}, \HERWIG~\cite{Corcella:2001bw}, and
\ISAJET~\cite{Baer:1999sp}, have enjoyed widespread use by
experimentalists.  These programs begin with a leading order hard
subprocess. Higher order effects are added by evolving the event using
the parton shower,
which provides an all orders description of parton emissions valid in
the soft and collinear regions, but is not accurate for well-separated
particles.  The partons are then grouped together into colorless
hadrons, resonances are decayed, and the underlying structure of the
event (beam remnants, multiple interactions, etc.) is added. These
programs are able to describe the exclusive structure of the event,
and so provide a useful input for subsequent detector simulation.

\NLOa\ integration programs go one order beyond
in the prediction of the cross section, have a reduced dependence on
arbitrary scale choices, and provide a good description of hard
well-separated emissions. They are able to predict distributions of
events, but are unable to produce events with the frequency predicted
by the theory (because the cancellations between Feynman diagrams are
usually achieved by allowing a fraction of phase space points
to have negative probability). These programs are commonly used by
experimental collaborations for determining $k$-factors, which are
often assumed to be constant, and are employed to correct the
normalization of distributions from leading order event generators.
Lacking individual events to evolve further, it is difficult to add
subsequent event features like hadronization or the underlying event.
This has meant that these programs are excellent theoretical tools for
predictions of distributions, but limits their usefulness for
producing events to be simulated in the detector environment.
The primary aim of the method presented in this paper is to improve
the usefulness of \NLOa\ calculations for experimental applications by
interpreting the result in a manner which is well suited for interface
to showering and hadronization generators and subsequent detector
simulation.

The pursuit of techniques for combining \NLOa\ calculations with the
parton shower is a natural direction for the evolution of event
generators.
Two primary challenges stand in the way: overlaps between phase space
volumes of differing dimensionality need to be accounted for in a
manner which does not double-count or neglect any region, and the
result needs to be interpretable in a probabilistic way (probabilities
should be everywhere positive definite).
Significant advances towards the resolution of the former challenge
have been achieved by Collins~\cite{Collins:2001fm} using a
subtractive approach, however negative weighted events are still
produced.


The issue of unweighted event generation has been addressed by the
author for the special case of diboson production in
Ref.~\cite{Dobbs:2001bx}, then further developed to include a
consistent merging of the parton shower in Ref.~\cite{Dobbs:2001gb}.
In the present study, these techniques are generalized and a veto
method proposed by P\"otter~\cite{Potter:2001an} is incorporated.
This allows for the organization of \NLOa\ event generators in an
elegant and simple manner which competes with leading order event
generators in terms of efficiency and computer time.

To illustrate the methods discussed in this paper, an event generator
for the process $\ppORppbar \rightarrow l^+l^- + X$ is constructed and
used to generate event distributions relevant to the Fermilab Tevatron
Collider and the CERN Large Hadron Collider (LHC). A precise knowledge
of the Drell-Yan lepton pair production process is particularly
important at hadron colliders. It will be used to probe new physics
(e.g.\ large extra dimensions, extra neutral gauge bosons), perform
precision measurements of electroweak parameters, constrain the parton
density functions, and calibrate the detector.  The latter is perhaps
most important to the physics program, because it means our knowledge
of this process will feed into the systematic errors for most physics
measurements: the lepton energy and momentum scale may be calibrated
{\it in situ} with $Z^0\rightarrow l^+ l^-$ events, the jet energy
scale may be determined using events with a $Z^0$ decaying to leptons
recoiling against a high transverse momentum jet, and the Drell-Yan
event rate may be used to determine the absolute luminosity. The
\NLOa\ QCD corrections to the process have been available for some
time~\cite{Altarelli:1979ub}.  Recently the complete \order(\alphaQED)
corrections have been calculated~\cite{Baur:2001ze}, and will play an
important role for precision measurements.

In the following section, background information relevant to the
\NLOa\ calculation is presented. A description of the phase space veto
method, numerical results, and shower evolution are presented in the
next sections.

%
%
\section{BACKGROUND} 

The \NLOa\ cross section receives contributions from the square of the
Born graphs, the interference of the Born graphs with the one-loop
graphs, and the square of the real emission graphs which contain an extra
colored parton in the final state,
\begin{equation}\label{e_matrixSq}
\Matrix^2_\txt{NLO} ~=~ \Matrix^2_\txt{Born} 
	+ \Matrix_\txt{Born} \otimes \Matrix_\txt{one~loop} 
	+ \Matrix^2_\txt{real~emission}.
\end{equation}
The second and third terms of Eq.~\ref{e_matrixSq} diverge when
treated separately, and so numerical integration calculations 
employ a regularization scheme which effectively combines pieces of
these terms to obtain finite results everywhere in phase space.

Commonly used schemes include the ``subtraction
method''~\cite{Ellis:1981wv}, ``dipole method''~\cite{Catani:1997vz},
and ``phase space slicing'' (\PSS)
methods~\cite{Bergman1989,Harris:2001sx,Fabricius:1981sx,%
Giele:1992vf,Giele:1993dj}.
All methods give identical results when used appropriately. For the
purposes of this study, features of the \PSS\ methods are convenient.
Variations of the \PSS\ method include the ``two parameter
\PSS''~\cite{Bergman1989} (see Ref.~\cite{Harris:2001sx} for an
accessible review), ``one parameter \PSS''~\cite{Fabricius:1981sx},
and ``$\smin$-slicing''~\cite{Giele:1992vf,Giele:1993dj}.

\begin{figure}
\noindent
  \mbox{\epsfig{file=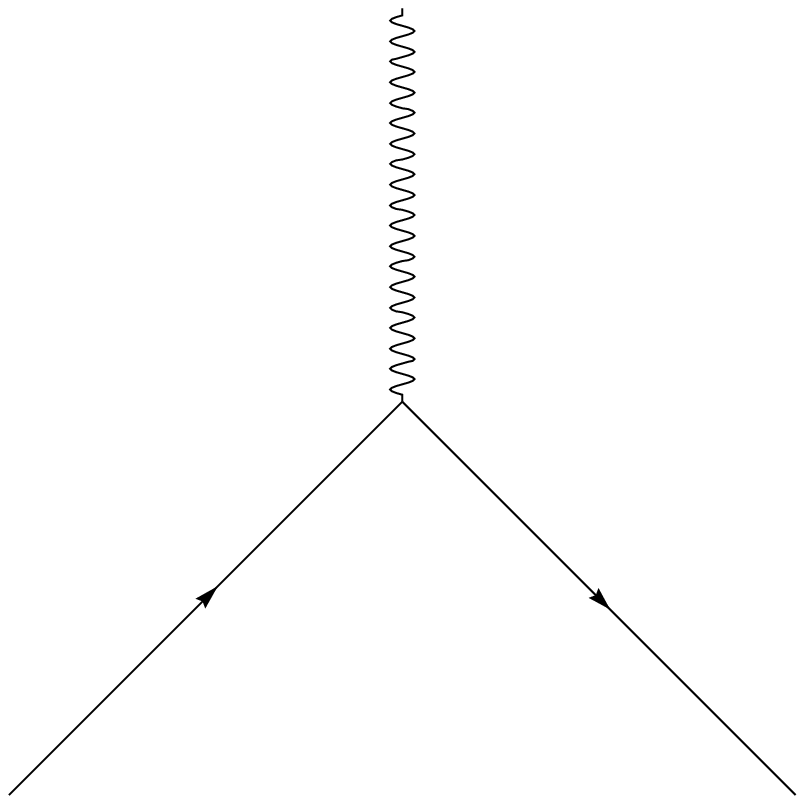,height=2.15cm}}
  \mbox{\epsfig{file=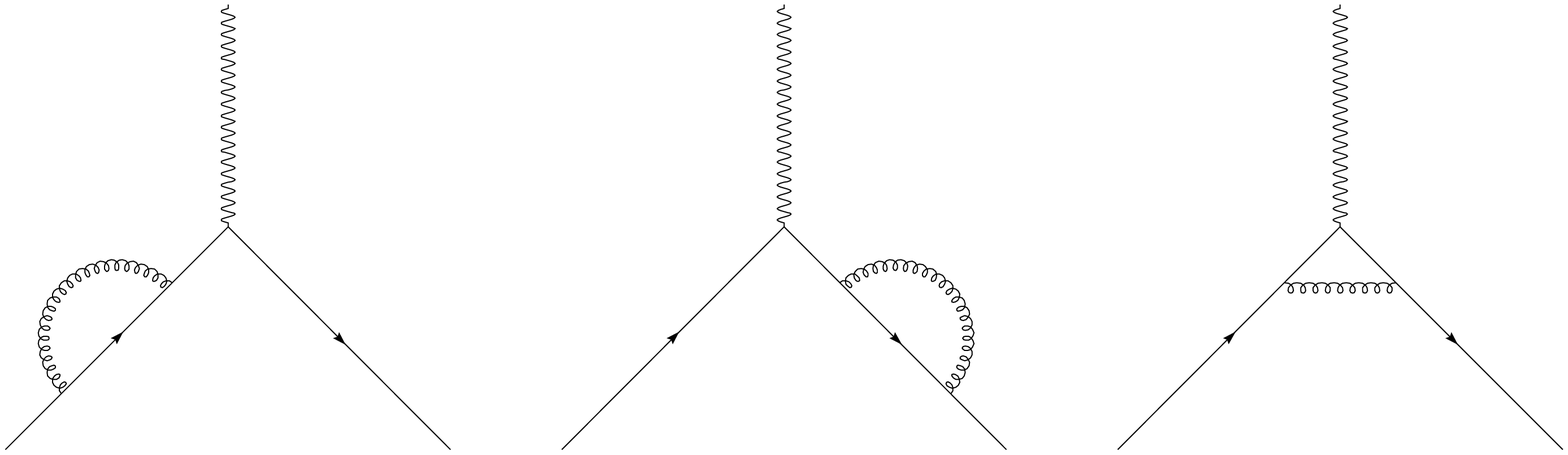,height=2.15cm}} \\
	{\footnotesize \mbox{} \hspace{.5cm} Born \hspace{3.25cm} one-loop } \\
  \mbox{\epsfig{file=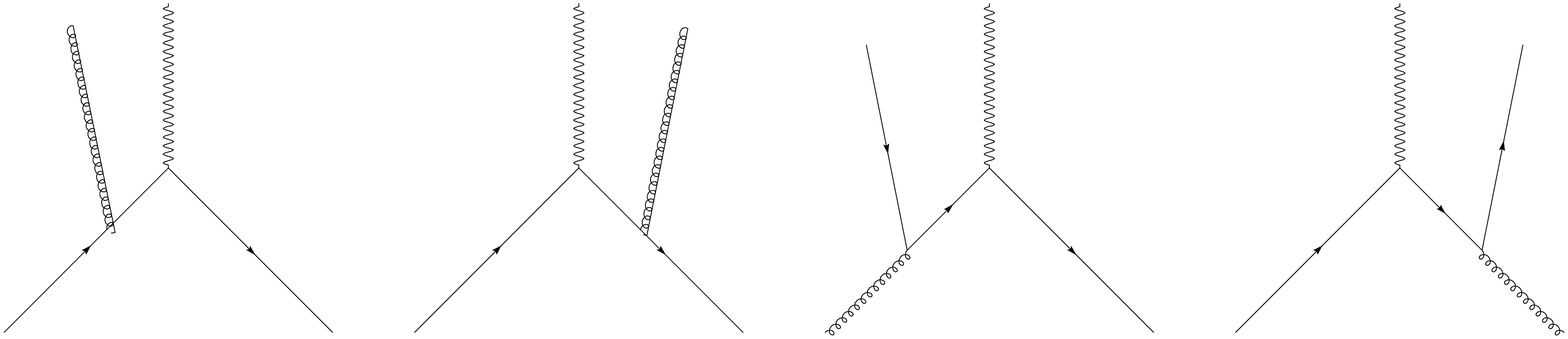,height=2.15cm}} \\
	{\footnotesize \mbox{} \hspace{3.5cm} real emission }

  \caption{ Feynman graphs contributing to $\ppORppbar\rightarrow
	Z^0+X$ at \NLOa.  The wavy line represents either a $Z^0$ or
	$\gamma^\star$, and the vector-boson decay products are not
	shown.  } \label{f_diagrams_Z_NLO}
\end{figure}

To illustrate the \PSS\ method, consider $\ppORppbar\rightarrow Z^0+X$
at \NLOa. The Feynman graphs are presented in
Figure~\ref{f_diagrams_Z_NLO}. The first term of Eq.~\ref{e_matrixSq}
is leading order (\alphaQED) and contains $n$ particles in the final
state. The phase space volume $\Phi_\txt{n}$ which defines the configuration
of the four-vectors is referred to as \nbody\ and specified by 4
degrees of freedom: the $Z^0$ mass, $Z^0$ boost, and two $Z^0$ decay
angles.  The second term of Eq.~\ref{e_matrixSq} is first order
(\alphaQED\alphaS) and is also described by \nbody\ kinematics.  The
third term of Eq.~\ref{e_matrixSq} is also first order
(\alphaQED\alphaS) and the final state contains the vector-boson (or
its decay products) and a colored parton (e.g.\ $Z^0 g,~Z^0 q,~Z^0
\bar{q}$) and is described by \nplusbody\ kinematics with 7 degrees of
freedom: the system mass, system boost, $Z^0$ mass, two $Z^0$
production angles, and two $Z^0$ decay angles.

For a particular choice of the \nbody\ kinematics, the phase space
$\Phi_{+1}$ which specifies the kinematics of the real emission is a
plane in $\hat{u}=(p_2-p_j)^2$~vs.~$\hat{t}=(p_1-p_j)^2$ space, shown
in Figure~\ref{f_veto-t_vs_u}, where $p_1$ and $p_2$ are the
four-momenta of the massless colliding partons, and $p_j$ is the
massless colored emission (the azimuthal degree of freedom is
unimportant and not shown).  The \nbody\ kinematics occupy a point at
the origin of this plane. The corresponding differential cross section
diverges as either axis of the plane is approached and at the origin,
i.e.\ when the emission becomes soft or collinear.

The \PSS\ methods regulate the singularities by partitioning the phase
space into a region of resolved emissions, and a region of unresolved
soft and collinear emissions. The resolved part is integrated
numerically.  The contribution from the unresolved soft and collinear
emissions is calculated analytically and included with the \nbody\
squared matrix element such that the net result is finite, though not
necessarily positive. For the case of $\smin$-slicing, the boundary of
the unresolved region is defined by a single parameter with dimension
energy squared. An emission is considered to be unresolved anytime the
invariant mass squared of any parton pair is less than the $\smin$
resolution parameter
\begin{equation}
	|s_{ij}|   
	~<~ \smin
	\hspace{1.5cm} \mbox{(unresolved region)}
\end{equation}
where the partons $i,~j$ may be either initial or final state.

The cross section for a particular \nbody\ configuration $\Phi_\txt{n}$,
integrated over the entire $\hat{u}$~vs.~$\hat{t}$ plane of
Figure~\ref{f_veto-t_vs_u} is a constant, schematically
\begin{equation}\label{e_smin_const}
\sigma^\txt{n}(\Phi_\txt{n},\smin) 
	~+~ \int_{s_{ij}>\smin}
	\sigma^\txt{n+1}(\Phi_\txt{n},\Phi_{+1})
	d \Phi_{+1}
	~=~ \const(\Phi_\txt{n})
\end{equation}
and is independent of the $\smin$ choice.  For a complete description
of $\smin$-slicing refer to~\cite{Giele:1992vf,Giele:1993dj}.

There exists a boundary in the unresolved region of
$\hat{t}$~vs.~$\hat{u}$ space, referred to here as $\szero$, inside
which the sum of the \nbody\ and \nplusbody\ contributions is
zero.  Knowing the location of the boundary, one could calculate the
\NLOa\ cross section and distributions by sampling only the
\nplusbody\ phase space, restricted to that region which lies above
the boundary (resolved partons). Thus
$\sigma^\txt{n}(\Phi_\txt{n},\szero)=0$ and the constant of
Eq.~\ref{e_smin_const} is
\begin{equation}
	\int_{s_{ij}>\szero} \sigma^\txt{n+1}(\Phi_\txt{n},\Phi_{+1})
	d \Phi_{+1}
	~=~ \const(\Phi_\txt{n}).
\end{equation}
The analytic expression for $\sigma^\txt{n}(\Phi_\txt{n},\smin)$ is given in
Ref.~\cite{Giele:1993dj}, and the analytic expression for $\szero$ is
derived in the appendix of the present paper.

A variant of the idea was originally proposed by Baer and Reno in
Ref.~\cite{Baer:1991qf}, who approximated the $\szero$ boundary as
constant and evaluated it by trial and error for single vector-boson
production in hadronic collisions using the two parameter \PSS\
method. 
However, the location of the boundary varies event-by-event with the
\nbody\ kinematics.  This was demonstrated by
P\"otter~\cite{Potter:2001an}, who formulated techniques 
for evaluating the $\szero$ boundary dynamically event-by-event. The
idea has been implemented for jet production in deep inelastic
scattering in Ref.~\cite{Potter:2001ej}, and good agreement is found
with the Hera data.


%
%
\section{ THE PHASE SPACE VETO METHOD } \label{s_pss_method}

To illustrate the phase space veto (\PSveto) method for hadronic
collisions, the process $\ppORppbar\rightarrow Z^0 + X \rightarrow l^+
l^- + X$ is chosen.  As for the \PSS\ methods, the phase space is
divided into two distinct volumes. The \nbody\ volume encompasses the
phase space with no resolved emission ($\ppORppbar\rightarrow Z^0
\rightarrow l^+ l^-$ kinematics), while the \nplusbody\ volume
describes the phase space with an extra parton in the final state,
$\ppORppbar\rightarrow Z^0 j\rightarrow l^+ l^- j$, where $j$ denotes
a gluon or (anti)quark.  For the PSS method, the two volumes would be
integrated separately using numerical techniques, and then added
together.

For the \PSveto\ method, the integration is organized differently.
Only the \nplusbody\ volume is integrated, and the \nbody\ matrix
elements are used to test on which side of the $\szero$ boundary each
phase space point lies.  An event candidate sampled in the
(unrestricted) \nplusbody\ phase space represents a point in the
$\hat{u}$~vs.~$\hat{t}$ plane shown in Figure~\ref{f_veto-t_vs_u}. If
the point lies below the $\szero$ boundary, the event is veto-ed. If
it lies above the boundary, it is assigned the event weight from the
\nplusbody\ differential cross section.  Since the location of the
$\szero$ boundary depends on both the factorization and
renormalization scales, the reduced scale dependence of the \NLOa\
calculation is maintained.

\begin{figure}
\noindent
\mbox{\epsfig{file=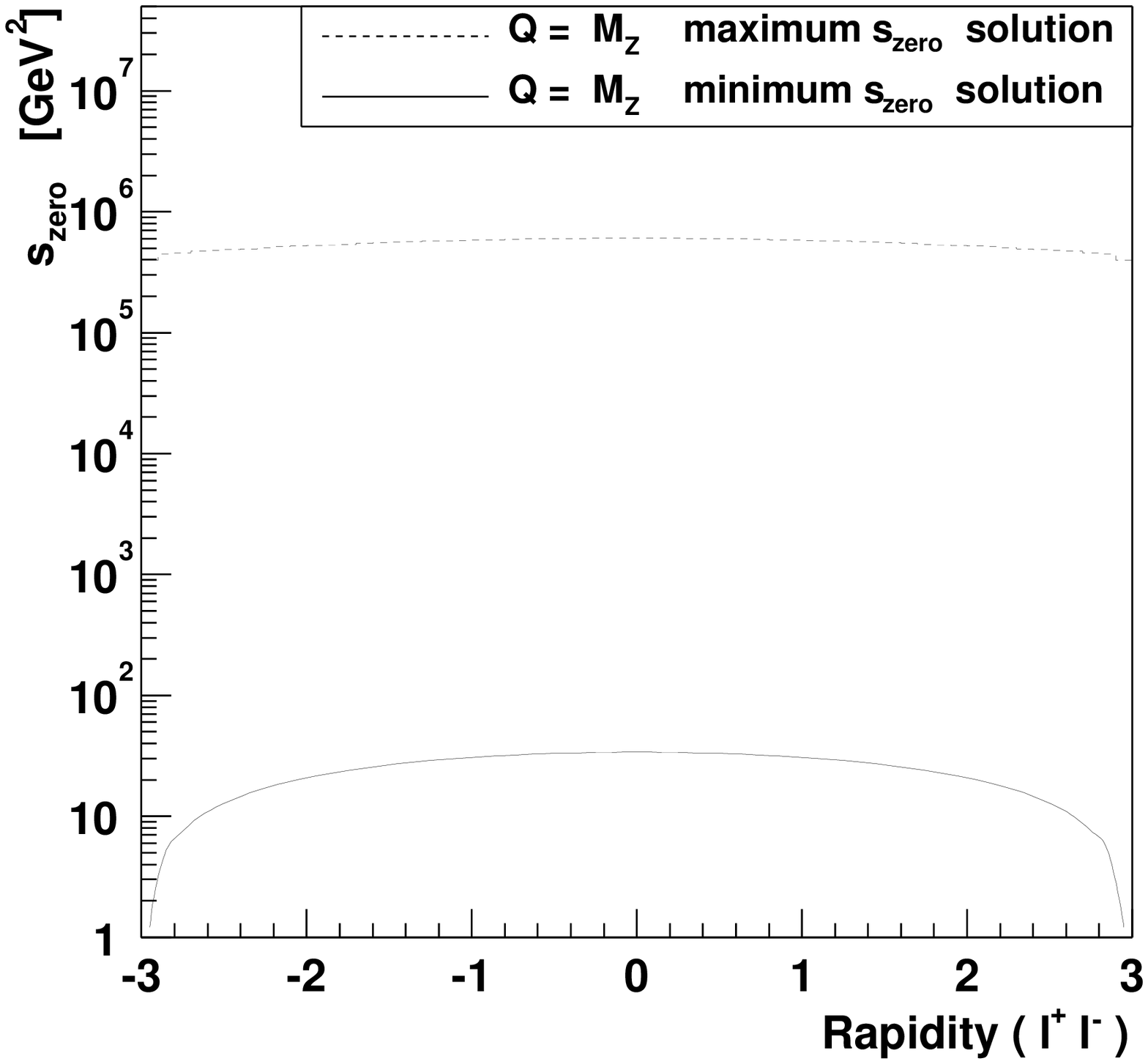,height=4.3cm,width=8.6cm}}
\\
\mbox{\epsfig{file=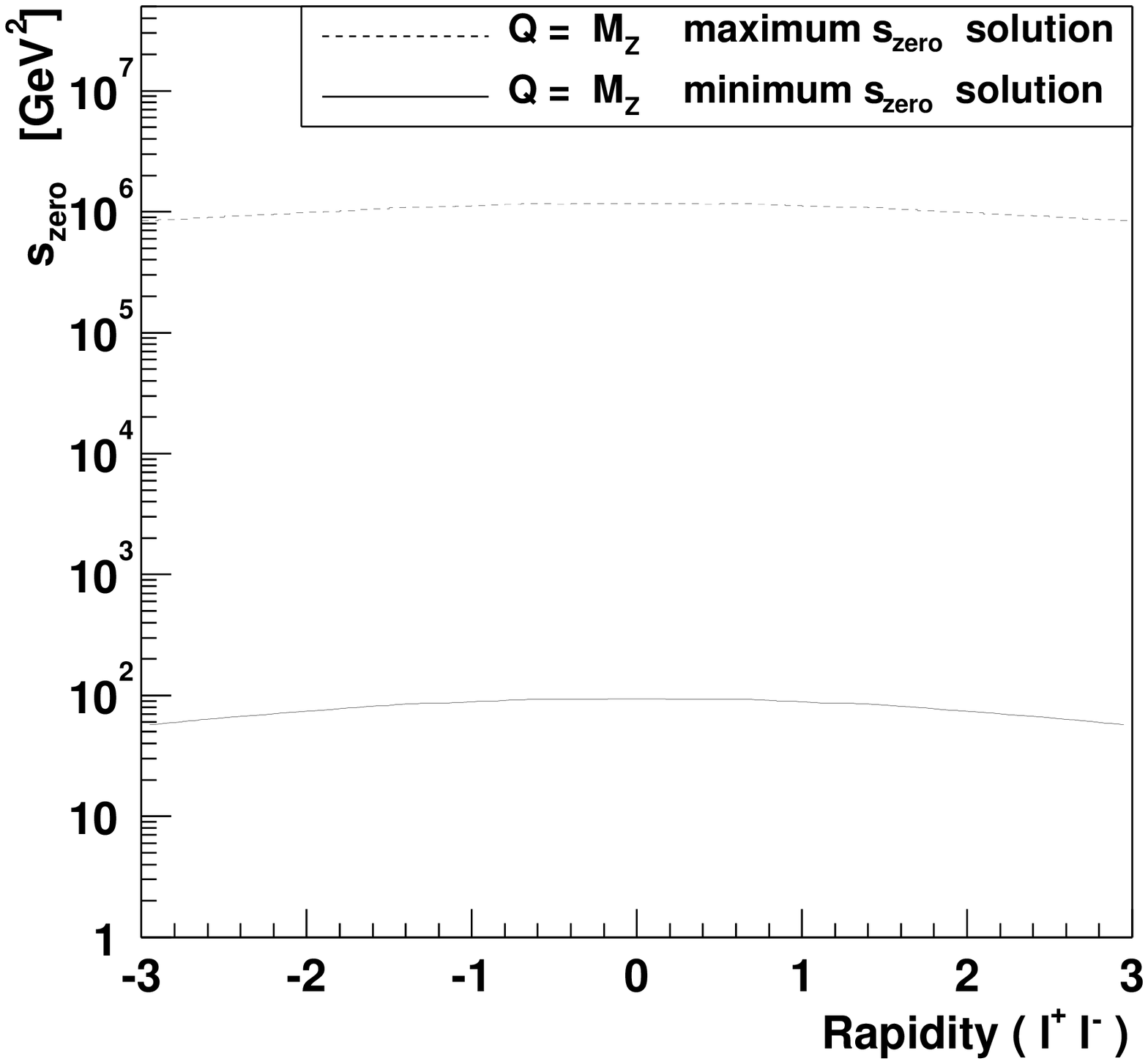,height=4.3cm,width=8.6cm}}
\caption{ The two roots of the quadratic \nbody\ differential cross
section presented in Eq.~\ref{e_xsec_nbody} are plotted as a function
of the lepton-pair rapidity, evaluated at parton center-of-mass energy equal
to the $Z^0$ mass for $p\bar{p}$ collisions at 2~TeV (Tevatron,
top) and for $pp$ collisions at 14~TeV (LHC, bottom). The smaller
solution is the $\szero$ function of interest, the larger solution
should not be interpreted physically.  }
\label{f_s_zero_solutions}
\end{figure}

There are at least two possibilities for determining on which side of
the boundary a phase space point lies:
\begin{enumerate}

\item the location of the $\szero$ boundary can be calculated
analytically.
The \nbody\ cross section is a quadratic equation in $\ln \smin$, with
the smaller of the two roots corresponding to the correct solution.
This is the method proposed in
Refs.~\cite{Potter:2001an,Potter:2001ej}, where the $\szero$ equations
for single jet production in electron-proton scattering are derived.
In the appendix of the present paper, the corresponding $\szero$
equations for $\ppORppbar\rightarrow Z^0/\gamma^\star$ at \NLOa\ are
derived.

\item without knowing the location of the $\szero$ boundary in the
unresolved region explicitly, it is possible to test on which side of
the boundary a phase space point lies by projecting the \nplusbody\
kinematics onto \nbody\ kinematics and simply evaluating the sign of
the \nbody\ matrix element with the $\smin$ boundary adjusted to sit
on top of the point in the $\hat{u}$~vs.~$\hat{t}$ plane. It is not
necessary to keep track of Jacobians from the projection nor overall
normalization factors, since only the sign of the matrix element is of
interest.  One must be careful
because well above the $\szero$ boundary (and after the
$\smin$-slicing approximation has broken down) the \nbody\ cross
section turns negative once again (corresponding to the second
solution of the quadratic equation in $\ln \smin$, discussed
above). In practice this happens only at large ($\approx 10^5 {\mathrm
GeV}^2$) values of $\smin \gg 100{\mathrm GeV}^2$ (see
Figure~\ref{f_s_zero_solutions}).  This strategy is simple to
implement, and works for $\smin$-slicing, one parameter \PSS, and two
parameter \PSS.\footnote{
   the two parameter \PSS\ method must first be expressed in terms of a
   single parameter, for example by defining $\deltaC=0.1\deltaS$ where
   $\deltaC$ and $\deltaS$ are the collinear and soft parameters of the
   method. The author has tested this for $p\bar{p}\rightarrow Z^0+X$ at
   \NLOa\ and found good agreement both with the unaltered two parameter
   \PSS\ method, and also with the \PSveto\ distributions presented in
   this paper.}
In this manner, processes which have already been coded as a numerical
integration using one of the \PSS\ methods can be re-cast as event generators
with minimal effort.
\end{enumerate}

Regardless of which of the above techniques is chosen, it is necessary
to project 7 dimension \nplusbody\ kinematics onto the 4 dimension
\nbody\ ones. This is accomplished by requiring the lepton-pair mass
$\mass{l^+l^-}$, and rapidity $Y_{l^+l^-}$, to remain unchanged in the
projection.
To perform the projection, the center of mass frame lepton momenta are
boosted into the vector-boson rest frame (which is the `new center of
mass frame'), and then boosted longitudinally such that the pair
regains their original rapidity, $Y_{l^+l^-}$.

\begin{figure}
\noindent
\mbox{\epsfig{file=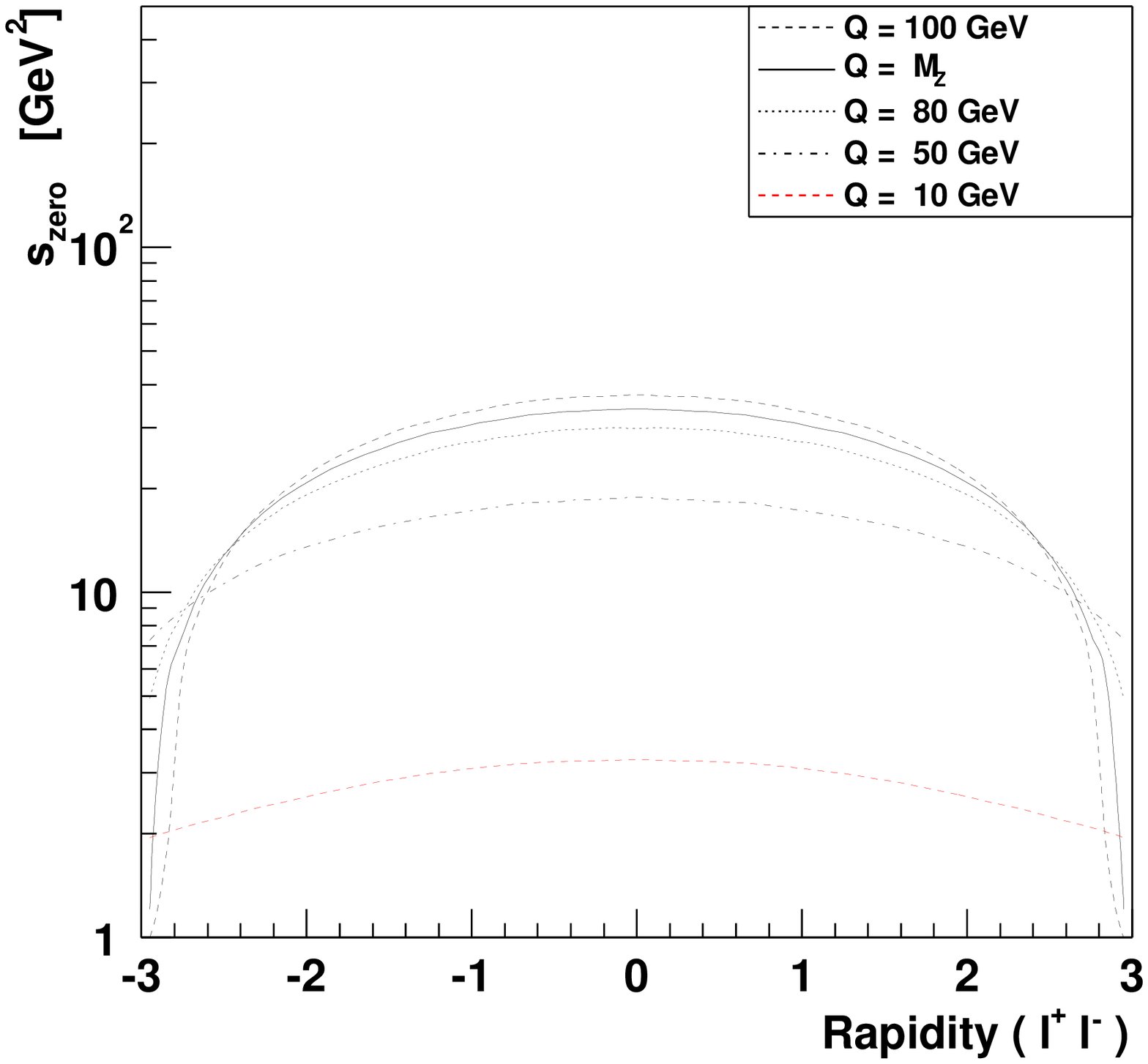,height=4.3cm,width=8.6cm}} \\
\mbox{\epsfig{file=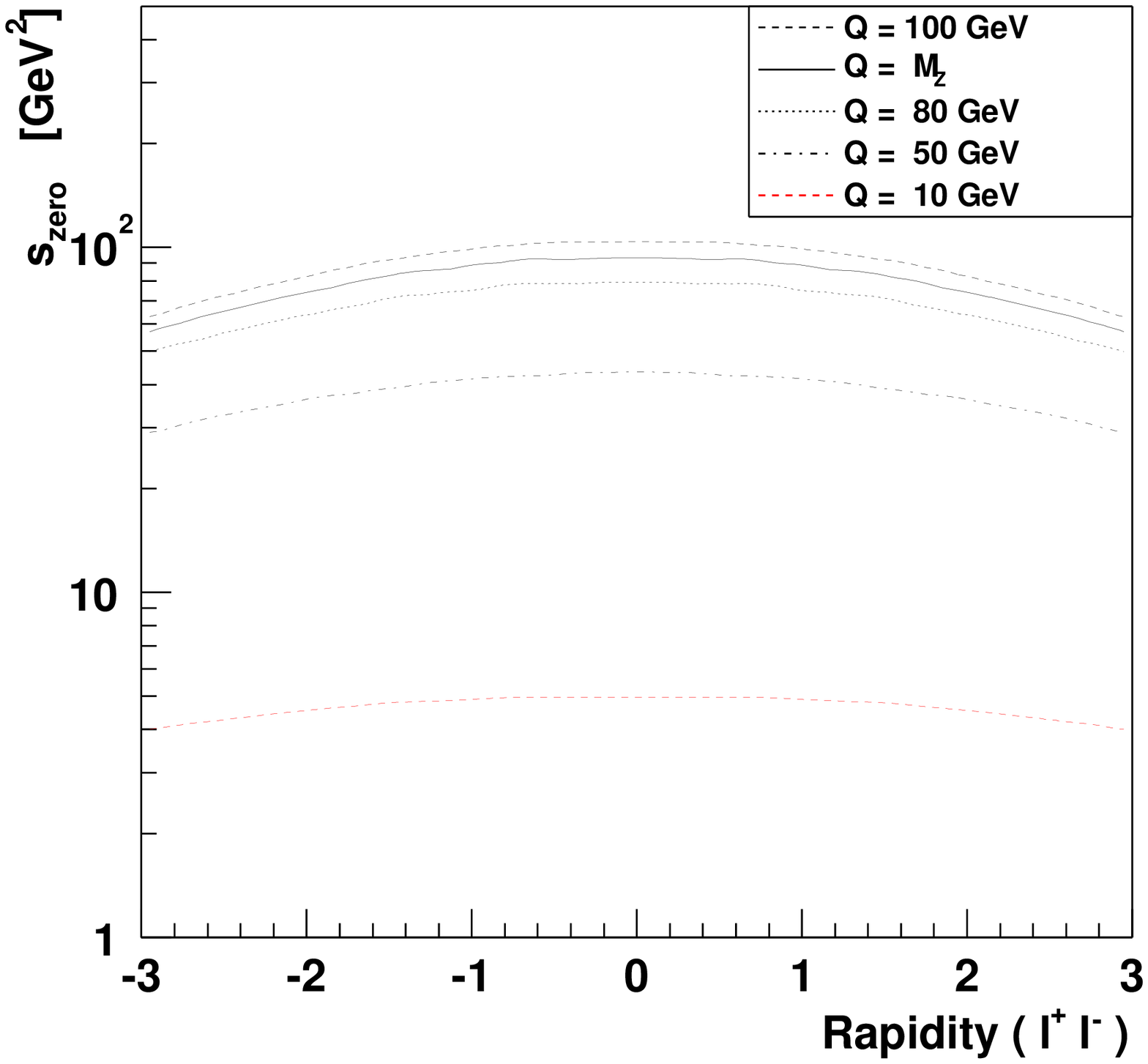,height=4.3cm,width=8.6cm}} 
  \caption{ The dependence of $\szero$ as a function of lepton-pair
  rapidity at several choices of parton the center-of-mass energy $Q$ is
  shown for the $p\bar{p}$ collisions at 2~TeV (Tevatron, top) and for
  $pp$ collisions at 14~TeV (LHC, bottom). The $\szero$ function does
  not depend strongly on the vector-boson decay angles.  }
\label{f_s_zero_vs_Q_LHC}
\end{figure}

\begin{figure}
\noindent
\mbox{\epsfig{file=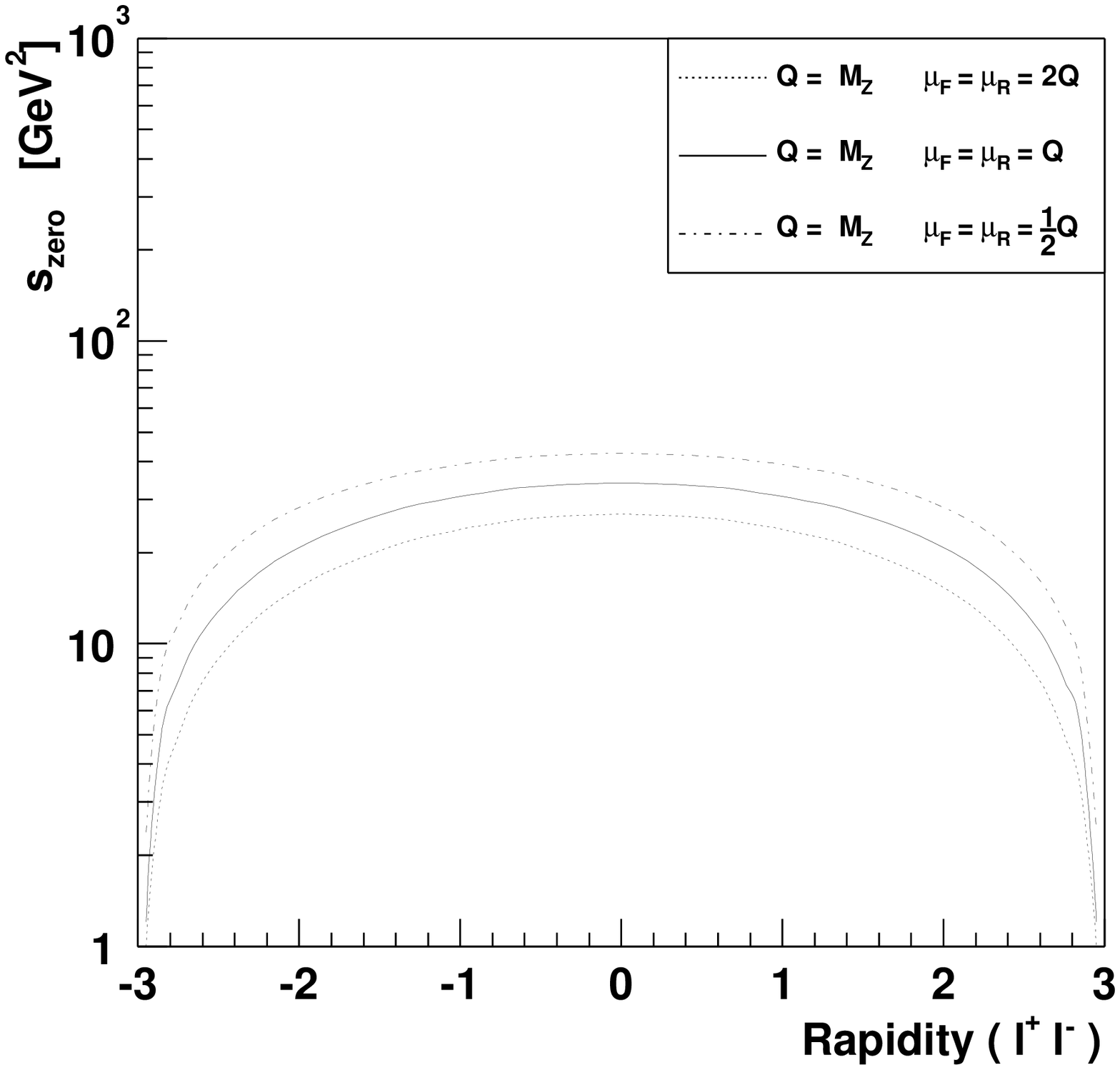,%
height=4.3cm,width=8.6cm}} \\
\mbox{\epsfig{file=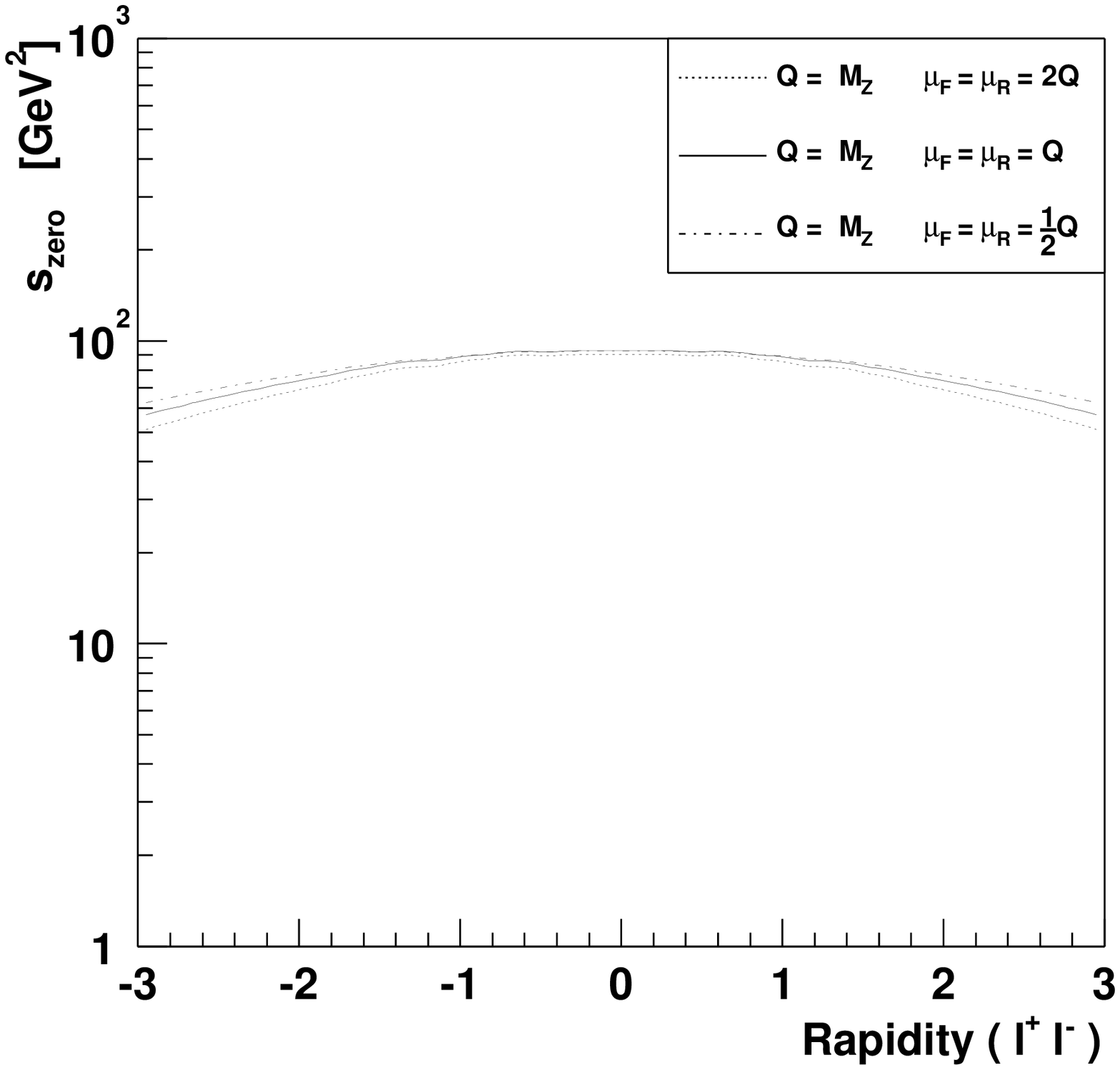,%
height=4.3cm,width=8.6cm}} 
\caption{ The scale variation of the $\szero$ function evaluated at
parton center-of-mass energy equal to the $Z^0$ mass is shown for
$p\bar{p}$ collisions at 2~TeV (Tevatron, top) and for $pp$ collisions
at 14~TeV (LHC, bottom). The $\szero$ function encodes information
about the factorization and renormalization scale choices into the
\PSveto\ method, preserving the \NLOa\ calculation's reduced scale
dependence.  }
\label{f_s_zero_scale_variation}
\end{figure}

In Figure~\ref{f_s_zero_vs_Q_LHC} the $\szero$ boundary for the
Tevatron and LHC collider energies are shown as a function of the
lepton-pair rapidity for several parton center of mass choices.  The
dependence of the $\szero$ boundary on the choice of renormalization
and factorization scales is shown in
Figure~\ref{f_s_zero_scale_variation}.

Though the $\szero$ boundary always exists, there is no
guarantee that the boundary lies within the region of validity for the
\PSS\ methods. This has not been a problem for the limited set of
processes to which the method has been applied. However, a hybrid of
the \PSS\ and subtraction methods~\cite{Glover:1995vz} has been proposed in
Ref.~\cite{Potter:2001an} to deal with the situation, should the need
arise.

 
%

For each phase space sample in the above algorithm, both the
\nplusbody\ and \nbody\ matrix elements are evaluated. This means that
the event generation will be slower than that for tree level events by
the amount of computer time it takes to evaluate the \nbody\ matrix
elements which are used to perform the veto.  Though this appears to
be the minimal computation necessary for performing a calculation
which incorporates the full \NLOa\ information, this is not the case.
There are ways in which the performance, in terms of computational
time, can be improved:
\begin{itemize}

\item upper and lower limits on $\szero$ can be evaluated
(see Figure~\ref{f_s_zero_vs_Q_LHC}).  For phase
space points which lie outside of these limits, the \nbody\ matrix
element need not be evaluated to determine whether or not the point is
veto-ed. For Tevatron energy, $\szero$ ranges from about 1~GeV$^2$ to
about 100~GeV$^2$.

\item since event generation is normally implemented using the
hit-and-miss (i.e.\ acceptance/rejection) Monte Carlo technique, the
majority of event candidates will be rejected anyway. The \PSveto\
need be applied only to those event candidates which are accepted (or
whenever an event candidate violates the maximum event weight against
which the acceptance/rejection is taking place). Since the efficiency
of event generators is typically about 25\% or lower, it means that
the \nbody\ matrix element needs to be evaluated rarely.  Further,
when the event candidates are sampled from an adaptive integration
grid (such as for the implementation presented in this paper), the
adaptive integration will ``learn'' the location of the boundary, and
will bias the sampling away from the region below the boundary.

\end{itemize}

\section{ NUMERICAL RESULTS } \label{s_numerical_results}

The $\ppORppbar\rightarrow Z^0 + X \rightarrow l^+ l^- + X$ event
generator is implemented using the squared matrix elements of
Ref.~\cite{Aurenche:1981tp} cast into the $\smin$-slicing
method~\cite{Giele:1993dj} (which employs special `crossed' structure
functions). The matrix elements include both the $Z^0$ and
$\gamma^\star$ diagrams with decay to massless leptons, such that the
branching ratio to one lepton flavor is automatically included. This
means finite width effects, lepton decay correlations, and
forward-backward asymmetries are everywhere taken into account.  The
generator is written in \CXX\ using modern object-oriented design
patterns. A new prototype \CXX\ version of the Bases/Spring
program~\cite{Kawabata:1995th} is used for adaptive integration and
event generation. Special care has been taken to make the program user
friendly,
and it is available upon request from the author.

All of the distributions and cross sections presented in this paper
are for $p\bar{p}$ collisions at 2~TeV (Tevatron Run~II) or $pp$ 
collisions at 14~TeV (LHC), with
the lepton-pair mass restricted~\footnote{
Hence the vector-boson is denoted by $Z^0$, even
though the $\gamma^\star$ contribution is included.}
 to the range 66-116~GeV and decaying to $e^-e^+$.
CTEQ3M~\cite{Lai:1997mg} parton density functions are used (chosen
because the `crossed' versions of the structure
functions~\cite{Giele:1993dj} are readily available, though in
principle they can be tabulated for any structure function). For all
calculations the renormalization and factorization scales have been
set equal to the vector-boson mass, $\mu_R=\mu_F=\mass{l^+l^-}$, and
the $\MSbar$ factorization scheme is used.  The input parameters are
chosen to coincide with those in \PYTHIA\ 6.200: the $Z^0$ mass and
width are $M_Z=91.188$~GeV and $\Gamma_Z=2.47813$~GeV, the electroweak
mixing angle is $\sin^2\theta_W=0.232$, and the electroweak coupling
is \alphaQED$(M_Z)=1/128.8$. The two-loop $\MSbar$ expression for
$\alphaS$ is used with $\Lambda^{4,\MSbar}=0.239$~GeV. Using these
input parameters, the \PSveto\ event generator predicts
\XSECTESTRESULT\ for the inclusive cross section at Tevatron Run~II.

\begin{figure}
\noindent
  \mbox{\epsfig{file=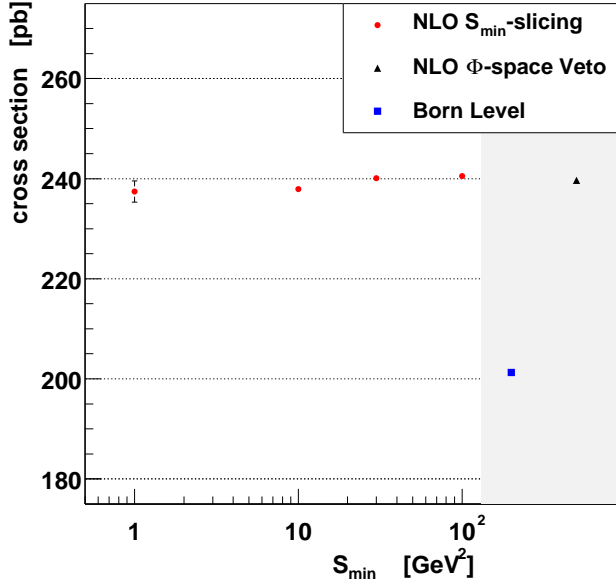,height=8.6cm}}
  \caption{ The inclusive \NLOa\ cross section for $p\bar{p}\rightarrow
  Z^0 + X \rightarrow e^+ e^- + X$ reactions at 2~TeV with the
  lepton-pair mass restricted to 66-116~GeV is shown as a function of
  the $\smin$ parameter for the $\smin$-slicing method. The cross
  section calculated using the \PSveto\ event generator is
  superimposed and is in good agreement.  The Born-level cross section
  is also shown. }
\label{f_xsec_vs_smin}
\end{figure}

In Figure~\ref{f_xsec_vs_smin} the inclusive cross section prediction
from the \PSveto\ event generator is compared to the predictions from
the $\smin$-slicing calculation using several choices of the $\smin$
parameter. The results are consistent, indicating the $\szero$
boundary lies within the region where the $\smin$-slicing
approximation is valid.

\begin{figure}
\noindent
\mbox{\epsfig{file=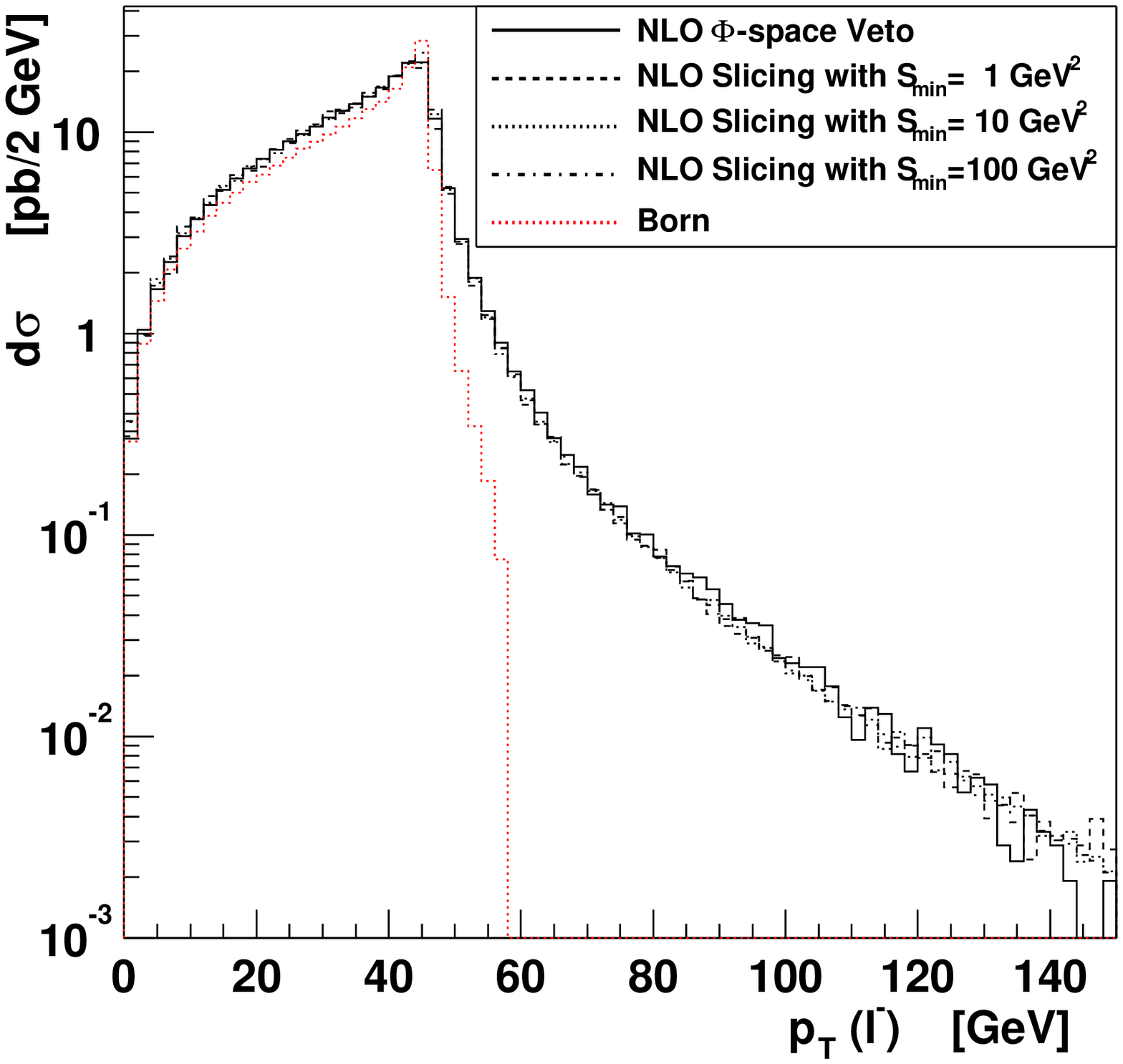,height=6cm,width=8.6cm}} \\
\mbox{\epsfig{file=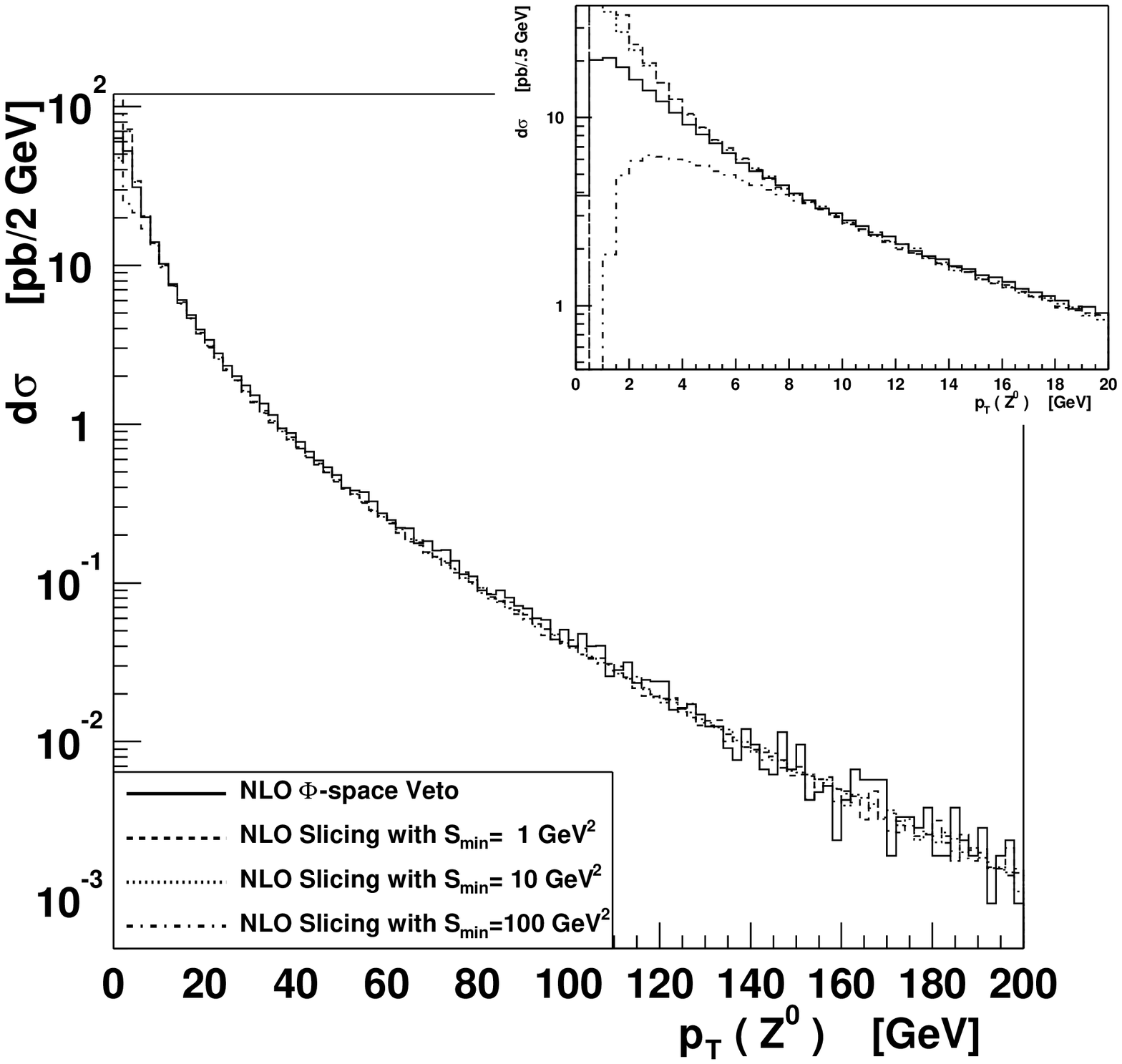,height=6cm,width=8.6cm}}
\caption{ The transverse momentum of the electron (top) and
vector-boson (bottom) are shown for the process $p\bar{p}\rightarrow
Z^0 + X \rightarrow e^+ e^- + X$ at 2~TeV with the lepton-pair mass
restricted to 66-116~GeV (no parton showering is used). Distributions
derived from numerical integrations at \NLOa\ using $\smin$-slicing
for various choices of the $\smin$ parameter are compared to the
distributions from the \NLOa\ \PSveto\ event generator. Agreement is
excellent everywhere, except in the low $\pT{Z}$ region (inset) where
fixed order perturbative QCD is unreliable.  The Born level prediction
is also super-imposed for the $\pT{e^-}$ distribution (top).  The Born
level prediction for the vector-boson transverse momentum is a delta
function at $\pT{Z}=0$.  }
\label{f_compare_to_smin}
\end{figure}

In Figure~\ref{f_compare_to_smin} distributions produced with the
\PSveto\ event generator are compared to those derived from numerical
integrations using $\smin$-slicing. The \PSveto\ method faithfully
reproduces the \NLOa\ transverse momentum of the electron. The
transverse momentum of the vector-boson also agrees well with the
$\smin$-slicing everywhere that the \NLOa\ calculation is valid.

\begin{figure}
\noindent
  \mbox{\epsfig{file=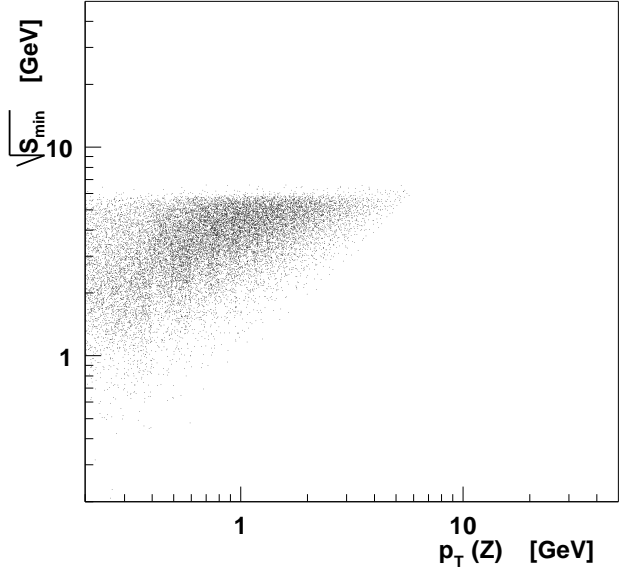,height=8.6cm,clip=}}
  \caption{ The kinematics of \PSveto\ method event candidates which
  have been veto-ed because they lie below the $\szero$ boundary are
  plotted in the $\pT{Z}$ vs. $\sqrt{\smin}$ plane. The largest
  $\pT{Z}$ of a veto-ed candidate event is 5.5~GeV.  The process is
  $p\bar{p}\rightarrow Z^0 + X \rightarrow l^+ l^- + X$ at 2~TeV with
  the lepton-pair mass restricted to 66-116~GeV. }
\label{f_SminVSpT}
\end{figure}

In the small $\pT{Z}$ region, multiple gluon emission becomes
important and fixed order perturbation theory is unreliable. This is
evident in the inset of Figure~\ref{f_compare_to_smin}. In this region
the results depend on the specific choice of the $\smin$
parameter. This is also the region where the \PSveto\ method becomes
unreliable because the minimum jet scale is coupled to the \nbody\
kinematics. This effect is visible in Figure~\ref{f_SminVSpT}, where
the kinematics of the veto-ed event candidates from the \PSveto\
method for a typical event generation run are plotted in the $\pT{Z}$
vs. $\smin$ plane. The largest $\pT{Z}$ of a veto-ed candidate event
is 5.5~GeV, indicating the \NLOa\ calculation is unable to provide a
useful prediction in the region below $\approx5.5$~GeV.  It is worth
stressing that this does not make the \PSveto\ method less useful than
$\smin$-slicing since any \NLOa\ calculation is unreliable here.  This
is the region where the distributions are better modeled with the
parton shower, and a suitable treatment which removes this minimum jet
scale coupling will be provided in the next section.  The $\szero$
boundary represents a lower limit to the usefulness of our fixed order
perturbative approximation. 
As such, $\szero$ is a useful concept as a qualitative measurement of
the frontier of the validity of our perturbative calculation.

\begin{figure}
\noindent
\mbox{\epsfig{file=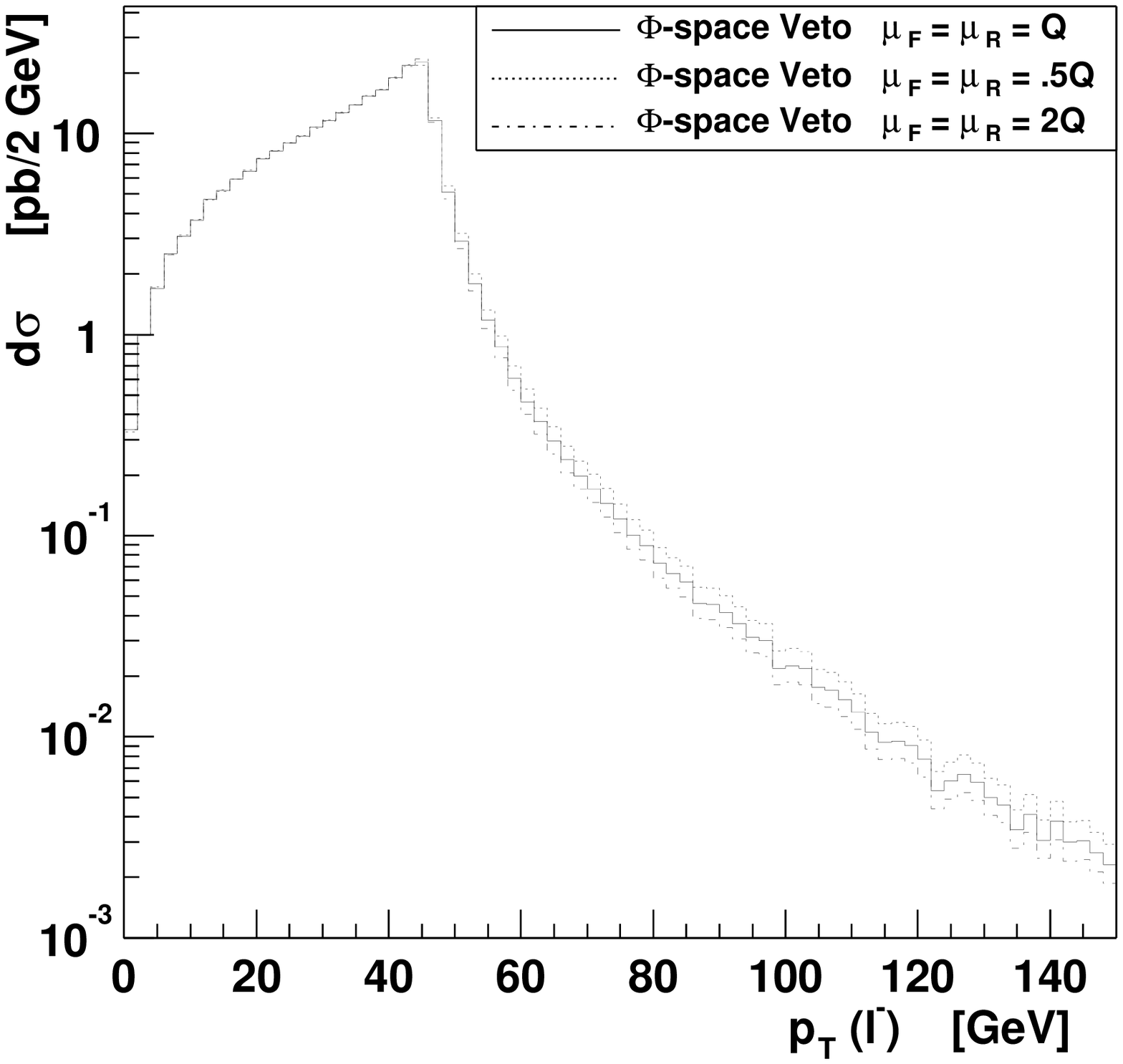,height=6cm,width=8.6cm}}
\mbox{\epsfig{file=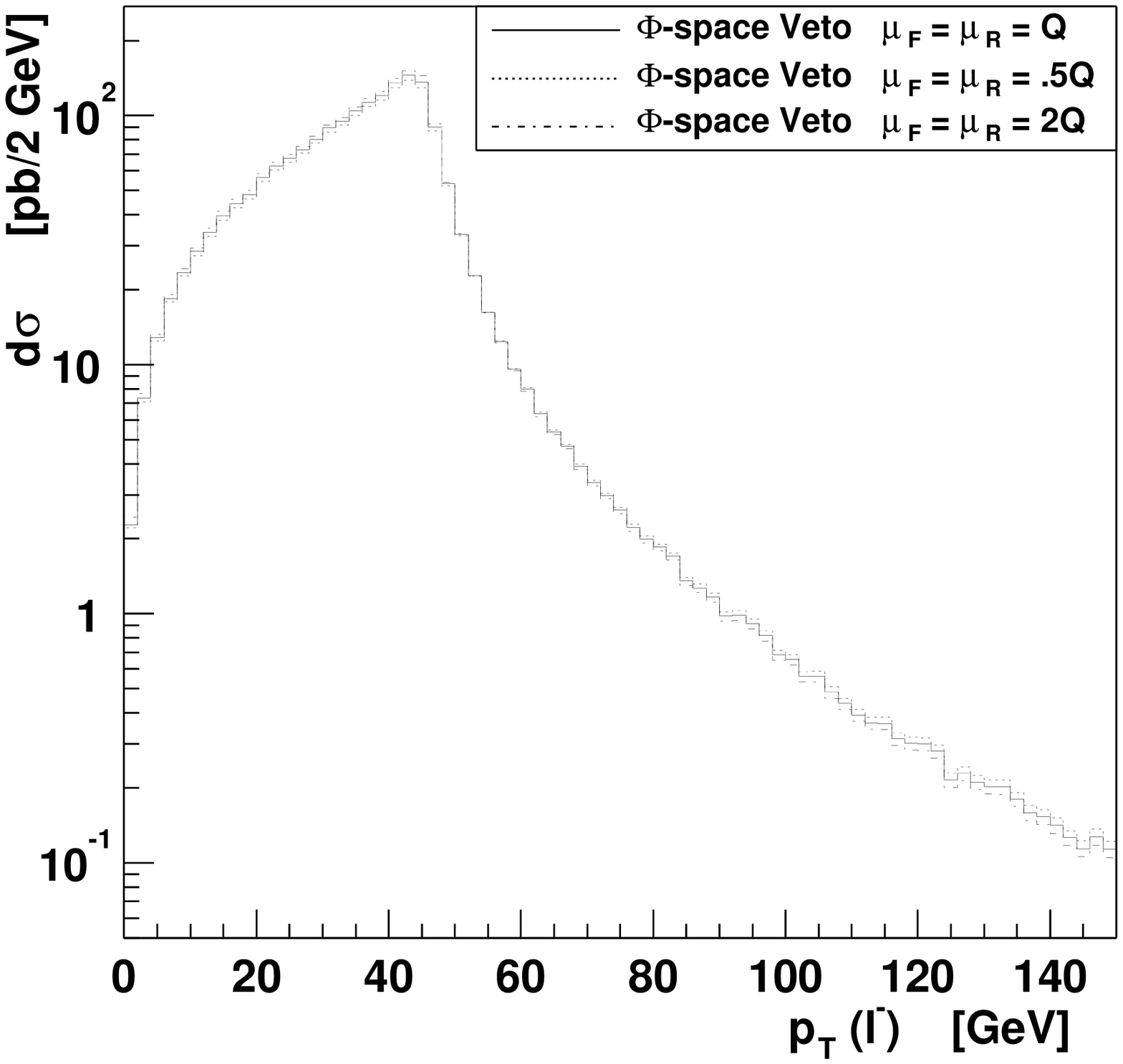,height=6cm,width=8.6cm}} \\
\mbox{\epsfig{file=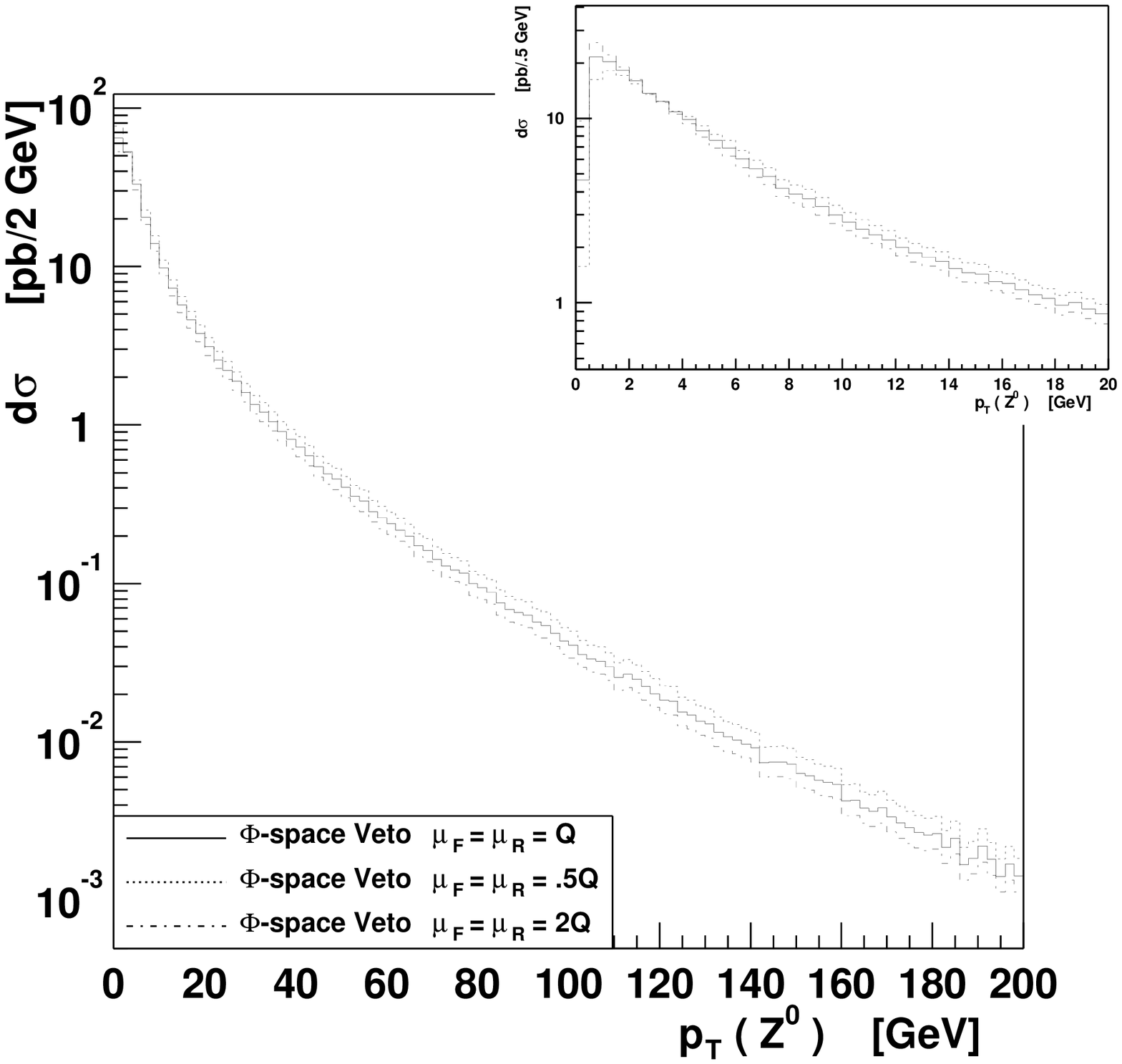,height=6cm,width=8.6cm}}
\mbox{\epsfig{file=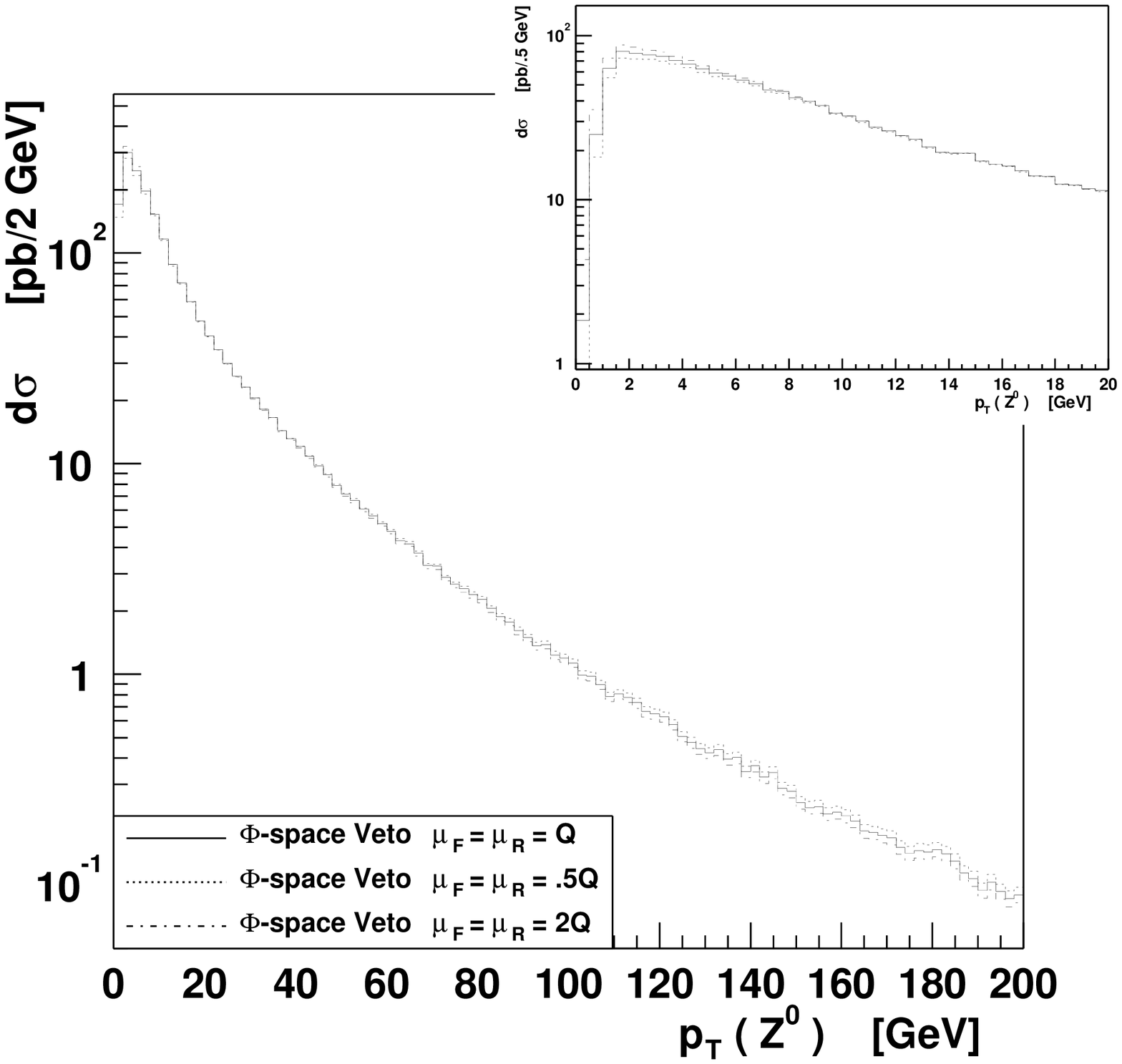,height=6cm,width=8.6cm}}
\caption{ The transverse momentum of the electron (top) and
vector-boson (bottom) are shown for the process $p\bar{p}\rightarrow
Z^0 + X \rightarrow e^+ e^- + X$ at \NLOa\ using the \PSveto\ method
(no parton showering is used) for different choices of the
renormalization and factorization scales. The spread in the
distributions is an indication of the theoretical error from neglected
higher order terms.  The distributions on the left are for 2~TeV
$p\bar{p}$ collisions at the Tevatron, and the ones on the right are
for 14~TeV $pp$ collisions at the LHC. The lepton-pair mass is
restricted to 66-116~GeV, and the three curves use the same event
sample.  }
\label{f_scale_variation}
\end{figure}

\begin{figure}
\noindent
\mbox{\epsfig{file=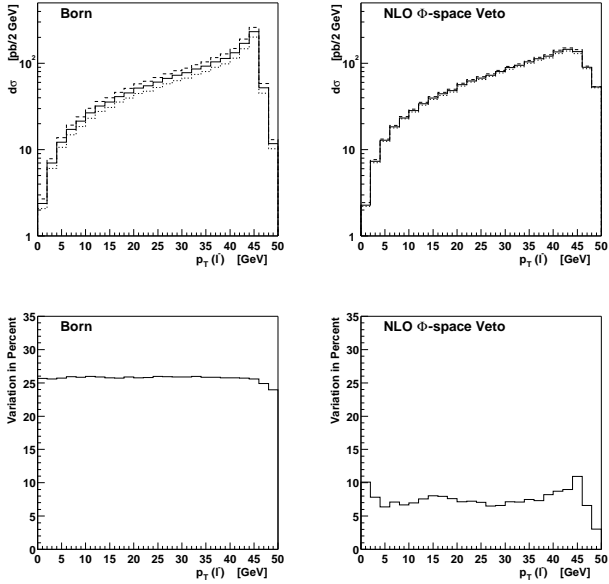,width=8.6cm}}
\caption{ The reduced scale dependence of the \NLOa\ \PSveto\
calculation as compared to the Born one is demonstrated.  The
transverse momentum of the electron (top) for the process
$pp\rightarrow Z^0 + X \rightarrow e^+ e^- + X$ with the lepton-pair
mass restricted to 66-116~GeV at LHC energy (14~TeV) is shown (the
vector-boson transverse momentum is not shown because the Born level
calculation does not provide a prediction for $\pT{Z}$).  The
renormalization and factorization scales are varied by a factor two in
the Born level calculation (left) and the \NLOa\ \PSveto\ calculation
(right).  The percent variation of the distributions is shown at
bottom. The variation is about a factor 4 smaller for the \NLOa\
\PSveto.
The effect is smaller at Tevatron energy in this region.  }
\label{f_compare_born_scale_variation}
\end{figure}

\begin{figure}
\noindent
\mbox{\epsfig{file=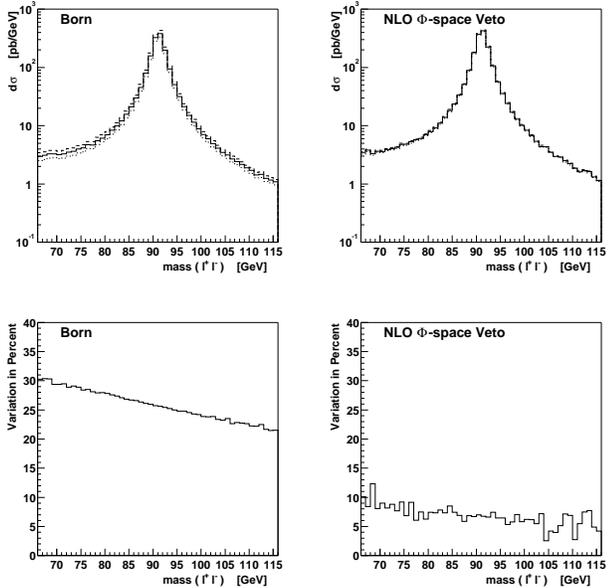,width=8.6cm}}
\caption{ The reduced scale dependence of the \NLOa\ \PSveto\
calculation as compared to the Born one is demonstrated.  The
lepton-pair mass in the vicinity of the $Z^0$ resonance is shown (top)
for the process $pp\rightarrow Z^0 + X \rightarrow e^+ e^- + X$ at LHC
energy (14~TeV).  The renormalization and factorization scales are
varied by a factor two in the Born level calculation (left) and the
\NLOa\ \PSveto\ calculation (right).  The percent variation of the
distributions is shown at bottom. The variation is about a factor 3
smaller for the \NLOa\ \PSveto. 
The effect is smaller at Tevatron energy.  }
\label{f_compare_born_scale_variation_mass}
\end{figure}

In Figure~\ref{f_scale_variation} the factorization and
renormalization scale dependence of the transverse momentum of the
electron and vector-boson distributions are shown using the \PSveto\
method for Tevatron and LHC energies. The scale dependence is
identical to that from the $\smin$-slicing method, because the
$\szero$ boundary encodes information about the scale choices
(Figure~\ref{f_s_zero_scale_variation}). The change in the
distributions resulting from the variation of the scales is an
indication of the theoretical error arising from neglected higher
order terms.  The importance of the reduced scale dependence is
demonstrated in Figure~\ref{f_compare_born_scale_variation}, where the
variation in the prediction at Born level and at \NLOa\ of the
transverse momentum of the electron for $pp\rightarrow Z^0 + X
\rightarrow e^+ e^- + X$ at LHC energy is shown. The comparison is
restricted to that region where the Born level prediction is
meaningful. The same comparison is shown for the lepton-pair mass
distribution in Figure~\ref{f_compare_born_scale_variation_mass}.  The
change in the scale at Born level results in more than a 25\%
variation in the distributions, whereas the prediction from the \NLOa\
\PSveto\ generator reduces this variation to about 7\%. The scale
dependence arising in predictions from event generators which use
leading order subprocesses (like \PYTHIA, \HERWIG, and \ISAJET) will
resemble that of the Born level prediction.


%
%
\section{ SHOWER EVOLUTION } \label{s_shower_evolution}

At the present stage, each event consists of the vector-boson decay
products and exactly one colored emission in the final state. The
energy scale of the emission is at least $\sqrt{\szero}$. Unweighted
events are provided by the Bases/Spring algorithm, and the
normalization is \NLOa. A coupling between the minimum emission scale
$\sqrt{\szero}$ and the kinematic configuration exists in the very
small $\pT{Z}$ region.

The next step is a consistent interface to a parton shower
algorithm. The goal is to have the parton shower dominate the
prediction in the soft/collinear region (in particular, it should
preserve the parton shower's prediction of Sudakov
suppression~\cite{Davies:1985sp}), and the first order tree level
diagrams dominate in the region of hard well separated partons. This
does not compromise the integrity of the prediction, it merely
highlights that different approaches are well-suited to different
regions.

To accomplish this, a parameter $\sPS$ is introduced to partition a
region of $\hat{t}$~vs.~$\hat{u}$ space which is exclusively the
domain of the parton shower. This parameter may be thought of as
separating the fixed order regime from the all-orders parton shower
region, in the same way that a \order(1 GeV) parameter in the
showering and hadronization programs defines the scale at which the
parton shower is terminated, and the simulation turns to the
non-perturbative hadronization model for a description of the physics.
This partition is shown in Figure~\ref{f_vetoPS-t_vs_u}.  Events which
lie below the $\sPS$ boundary are first projected onto \nbody\
kinematics (i.e. the point in $\hat{t}$~vs.~$\hat{u}$ space is moved
to the origin) and the parton shower is allowed to evolve the event
out into the plane. The projection is performed keeping the
lepton-pair mass and rapidity fixed, exactly as described in
Sec.~\ref{s_pss_method}. 
The parton shower is invoked with the scale set to $\sqrt{\sPS}$,
which ensures the evolution does not move the event out into a region
of phase space which has already been counted using the first order
tree level matrix elements.

A reasonable choice for the $\sqrt{\sPS}$ parameter is a few times the
minimum jet scale, $\sqrt{\szero}$. This ensures the first order tree
level matrix element is reliable above the $\sPS$ boundary. The
distributions have very little sensitivity to the choice of $\sPS$.

For events which lie above the $\sPS$ region, the parton shower
is also invoked, this time with a scale equal to the minimum invariant
mass of any parton-pair 
\begin{equation}
\mbox{parton~shower~scale} ~=~ \mbox{minimum} \left[~
	Q_{q\bar{q}},~ Q_{qg},~ Q_{\bar{q}g} ~\right]
\end{equation}
which ensures no double counting can occur.

\begin{figure}
\noindent
\mbox{\epsfig{file=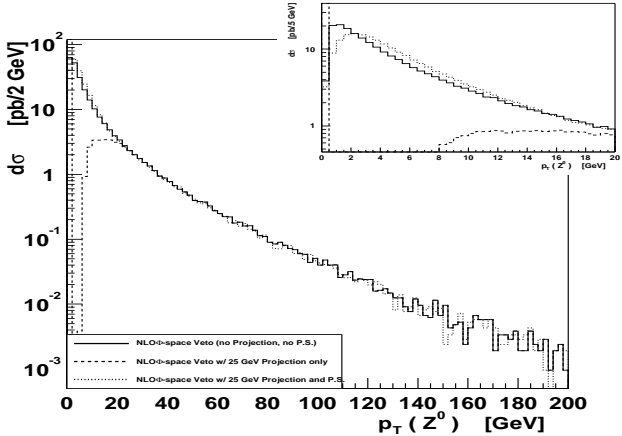,height=6cm,width=8.6cm}}
  \caption{ The $\pT{Z}$ distribution is shown after different stages
  of the event generation for $p\bar{p}\rightarrow Z^0 + X \rightarrow
  e^+ e^- + X$ at 2~TeV with the lepton-pair mass restricted to
  66-116~GeV. The solid line is the \PSveto\ \NLOa\ distribution
  without any projection or parton shower.  The dashed line is the
  (nonphysical) distribution for the same event sample, after applying
  the projection with $\sqrt{\sPS}=25$~GeV.  The dotted line is the
  distribution after subsequent evolution through the showering and
  hadronization program.  }
\label{f_see_projection}
\end{figure}

The effect of the projection and subsequent parton showering is shown
in Figure~\ref{f_see_projection}. Initially the distributions are
provided by the \PSveto, solid line. The projection is applied to
events which sit below the $\sPS$ boundary, which effects only the
small $\pT{Z}$ region, and is shown as a dashed line and does not
correspond to anything physical. Finally the parton shower is applied
(dotted line), and has the largest effect on those events which have
been projected.

\begin{figure}
\noindent
\mbox{\epsfig{file=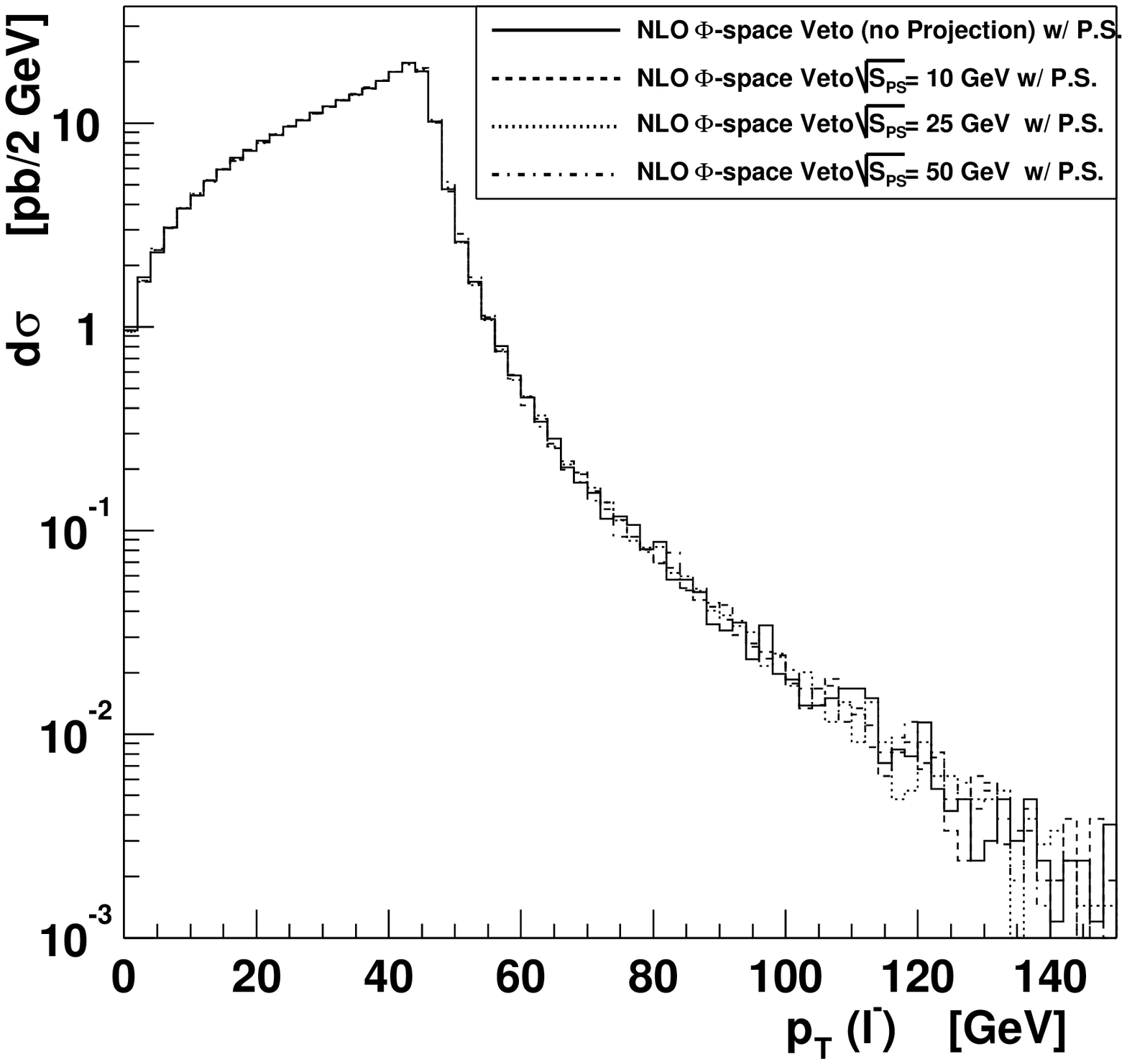,height=6cm,width=8.6cm}} \\
\mbox{\epsfig{file=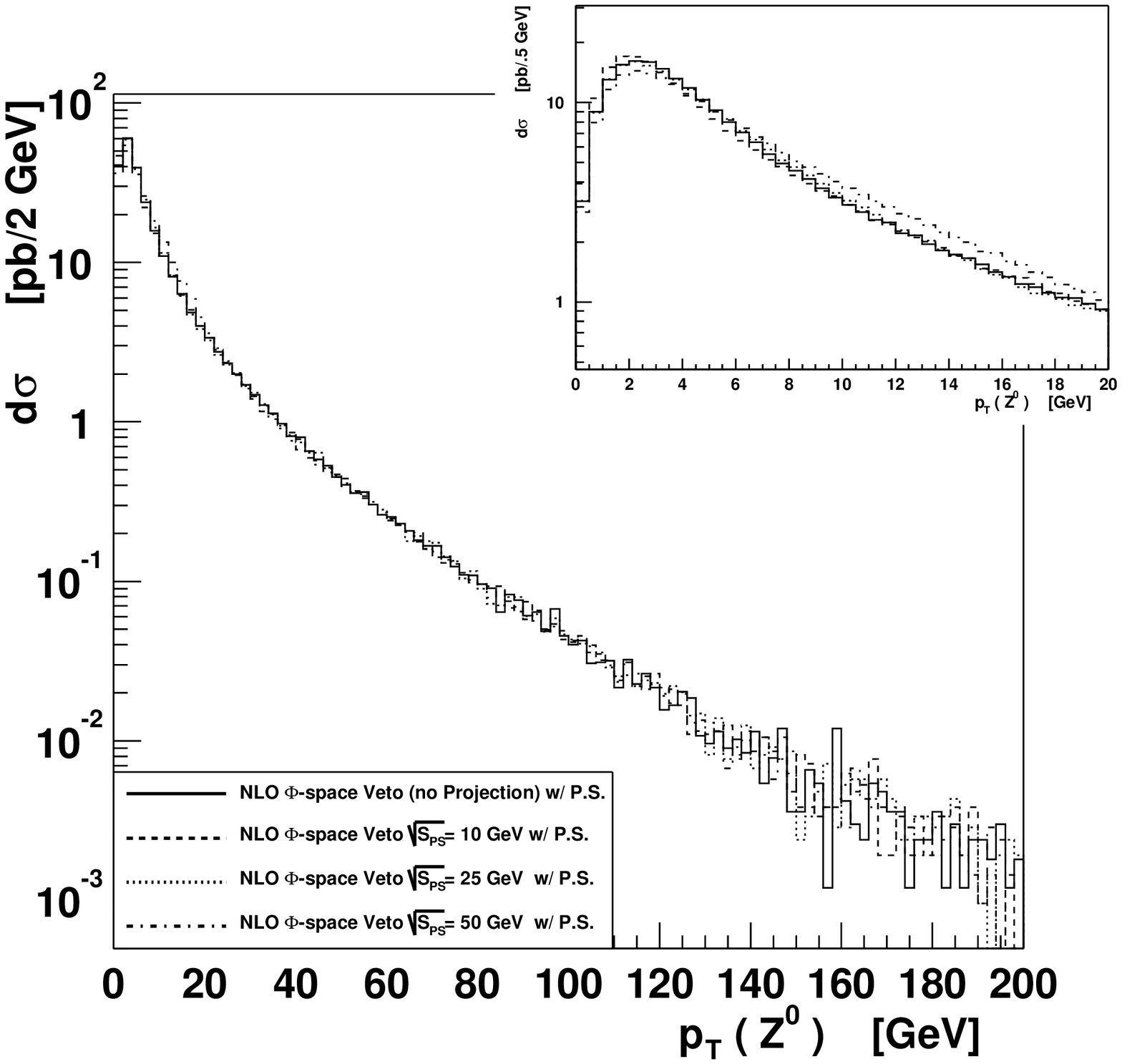,height=6cm,width=8.6cm}}
\caption{ The effect of the parton shower on the \PSveto\
distributions is shown for several choices of the $\sPS$ parameter
which partitions the phase space into the region populated by the
parton shower, and the region populated directly by the first order
matrix elements.  The transverse momentum of the electron (top) and
vector-boson (bottom) are plotted for the process $p\bar{p}\rightarrow
Z^0 + X \rightarrow e^+ e^- + X$ at 2~TeV with the lepton-pair mass
restricted to 66-116~GeV. There is very little dependence on the
specific choice of the $\sPS$ parameter. }
\label{f_projection}
\end{figure}

In Figure~\ref{f_projection} the \PSveto\ event distributions
(including parton shower evolution) are shown for several choices of
the $\sPS$ parameter. The dependence on the $\sPS$ parameter choice is
small, indicating discontinuities which might exist at the $\sPS$
boundary are also small.

For the distributions presented here, events from the \PSveto\
generator have been evolved with the \PYTHIA\ 6.200 parton shower.
\PYTHIA\ is attached using the {\tt HepUP}
interface~\cite{Boos:2001cv}, which is a generic standard for the
communication between event generators.
Having evolved the events through the parton shower, \PYTHIA\ provides
other features of the event structure such as hadronization,
resonance decays, beam remnants, and multiple interactions. The
showered event distributions presented in this paper include all of
these features.  The use of the {\tt HepUP} interface allows for the
parton shower program to be easily interchanged.  The choice of
\PYTHIA\ is arbitrary, there is nothing which precludes the use of any
other showering program.\footnote{
 For the case of the \HERWIG\ parton shower, there is a region or
 ``dead zone'' in the $\hat{t}$~vs.~$\hat{u}$ plane of
 Fig.~\ref{f_veto-t_vs_u} where emissions never occur.  The boundary
 of the dead zone is a natural choice for the partition which
 separates the parton shower region from the region populated directly
 by the first order matrix element when using \HERWIG. This is the
 prescription employed in Ref.~\cite{Corcella:1998rs} for ``hard matrix
 element corrections'' to single vector-boson production.  }

The full event generator is now complete. Adaptive integration and
phase space generation is provided by BasesSpring. The event weights
are evaluated using the \PSveto\ method, which discards those event
candidates lying below the $\szero$ boundary. When the program is
executed, the phase space is first mapped onto a grid using an
initialization pass with the adaptive integration (performed by the
`Bases' part of the BasesSpring package).  The `Spring' part of the
BasesSpring package then provides unweighted events, by sampling
candidate events from the adaptive integration grids and accepting
events according to the differential cross section using the
acceptance-rejection algorithm.
After removing the emission from those events which are soft or
collinear (as defined by the $\sPS$ parameter), the events are
transferred to the \PYTHIA\ package using the {\tt HepUP}
interface. \PYTHIA\ performs the parton shower, and subsequent event
evolution including hadronization, etc. 

While the end result in terms of physics does not differ significantly
from that obtained by the author in Ref.~\cite{Dobbs:2001gb} for $WZ$
production, the method presented here is simpler, easier to implement,
faster in terms of computer time, and may be generalized to a broad
range of processes.  Improved methods for invoking the parton shower
from parton level event configurations are being
developed~\cite{Catani:2001cc,Andre:1997vh}, and are suitable for
application to the \PSveto\ events. \footnote{
The {\tt apacic++}~\cite{Krauss:1999fc} showering program employed in
Ref.~\cite{Catani:2001cc} does not yet include initial state showers,
but an implementation is expected soon.}

A comparison of the computer time for generating the events is
presented in Table~\ref{t_computer_time}.  The processing time per
event and generation efficiency (percentage of candidate weighted
events which are accepted in the event generation algorithm) for
\PYTHIA\ and the \PSveto\ are similar, indicating the \PSveto\ method
is successful in encoding the extra \NLOa\ information without
affecting the overall time-performance of event generation.

\begin{figure}
\noindent
\mbox{\epsfig{file=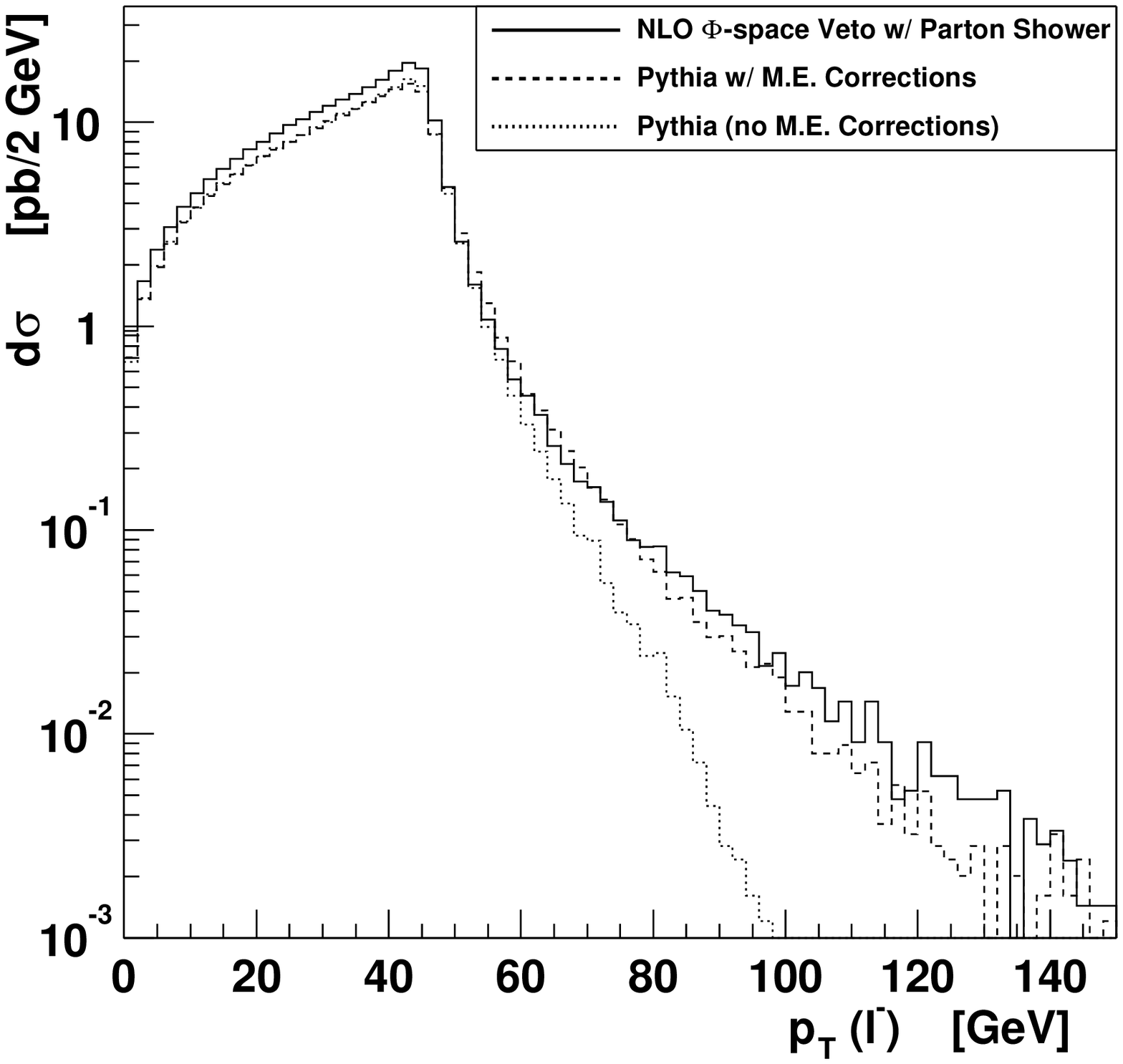,height=6cm,width=8.6cm}} \\
\mbox{\epsfig{file=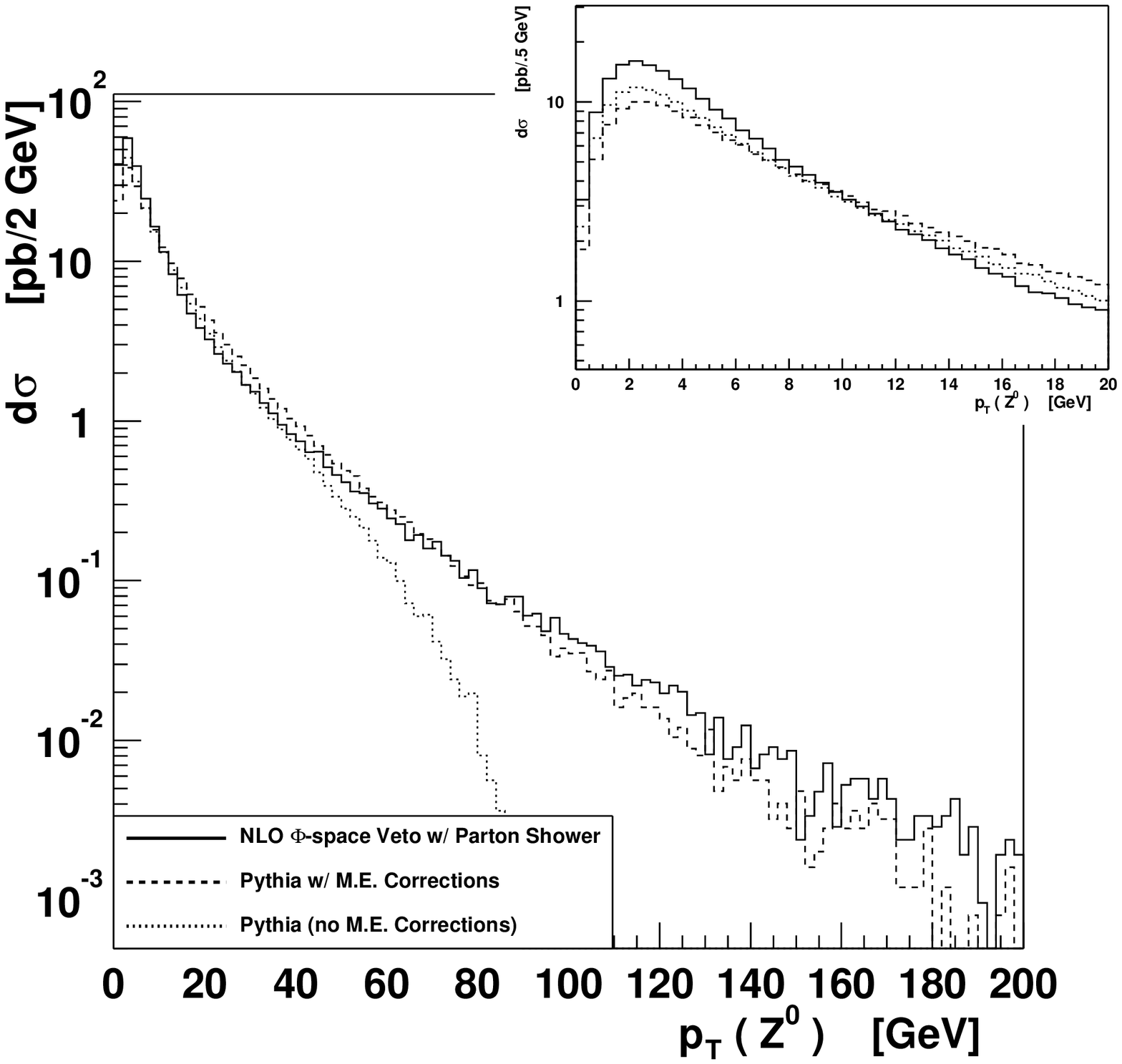,height=6cm,width=8.6cm}} 
  \caption{ Distributions for the process $p\bar{p}\rightarrow Z^0 + X
	\rightarrow e^+ e^- + X$ at 2~TeV from the \PSveto\ event
	generator (solid line, includes evolution through the \PYTHIA\
	shower and hadronization program) are compared with the
	\PYTHIA\ internal process distributions. The dashed line is
	the ``matrix element corrected'' \PYTHIA\ prediction and the
	dotted line is the ``old'' (no matrix element corrections)
	\PYTHIA\ prediction. The lepton-pair mass is restricted to
	66-116~GeV.  }
\label{f_compare_pythia}
\end{figure}

In Figure~\ref{f_compare_pythia} the \PSveto\ distributions (solid
line, includes evolution with the \PYTHIA\ showering and hadronization
package) are compared to the predictions from \PYTHIA.  In \PYTHIA\
there are two strategies implemented for single vector-boson
production. For both strategies the hard subprocess is chosen
according to the Born level matrix element, such that the
normalization is always leading order. For the ``old'' \PYTHIA\
implementation of the process, the event is then evolved with the
standard parton shower beginning at a scale equal to the vector-boson
mass. For the new ``matrix element (M.E.) corrected'' implementation
of the process,~\cite{Miu:1999ju} the shower is initiated at a scale
equal to the machine energy and is corrected according to the
$Z^0$+jet first order tree level matrix element, which results in a
considerable improvement of the high $\pT{Z}$ region modeling.  The
virtual one-loop contribution is not included anywhere in the \PYTHIA\
implementations.  The dotted line in Figure~\ref{f_compare_pythia} is
from the standard \PYTHIA\ process, and the dashed line is from the
M.E.\ \PYTHIA\ process.  The \PSveto\ distribution and M.E.\ corrected
\PYTHIA\ shapes are rather similar, indicating the matrix element
corrections in \PYTHIA\ are having the desired effect. The \PSveto\
distributions have the advantage of \NLOa\ normalization and a reduced
dependence on the factorization and renormalization scales.

%
%
\section{ CONCLUSIONS } \label{s_conclusions}

The \PSveto\ method for organizing \NLOa\ calculations into event
generators is demonstrated for $Z^0$ production in hadronic
collisions. The method is based loosely on the ideas proposed by
P\"otter for deep inelastic scattering~\cite{Potter:2001an}. The
primary motivation for the method is to move numerical \NLOa\
calculations beyond the status of ``event integrators'' to ``event
generators'', making them suitable for interface to showering and
hadronization programs and subsequent detector simulation.

The general features of the \PSveto\ method are:
\begin{itemize}
\item event weights are positive definite, meaning the standard
methods for event generation can be applied, providing a prediction
which is well suited for experimental applications.
\item in the soft/collinear region, the results are dominated by the
parton shower. In particular the low $\pT{}$ region exhibits
Sudakov suppression.
\item in the region of hard well separated partons, the distributions
are dominated by the first order matrix element.
\item the normalization is \NLOa\ and the reduced scale dependence
afforded by the \NLOa\ calculation is maintained.
\end{itemize}

The method has been implemented as an event generator (available from
the author) for $\ppORppbar\rightarrow Z^0/\gamma^\star + X
\rightarrow l^+ l^- + X$, with showering and hadronization provided by
the \PYTHIA\ package.

%
%
\section*{ ACKNOWLEDGMENTS } \label{s_acknowledgements}

The author would like to thank the ATLAS Collaboration and in
particular M.~Lefebvre and I.~Hinchliffe.  I am grateful to U.~Baur,
W.~Giele, and T.~Sj\"ostrand for informative discussions. I thank the
organizing committee of the {\it Physics at TeV Colliders Workshop},
May 21 -- June 1, 2001 in Les Houches, France. I am indebted to S.\
Kawabata for allowing me use of a prototype version of the
Bases/Spring program.
This work has been supported by the Natural Sciences and Engineering
Research Council of Canada.

%
%
\section*{ APPENDIX: The $\szero$ Function for $\ppORppbar \rightarrow Z^0
+ X$ at \NLOa }

The differential cross section for the \nbody\ contribution to
 $\ppORppbar \rightarrow Z^0 + X$ evaluated in the $\MSbar$ scheme,
 integrated over unresolved emissions out to a scale of $\smin$, and
 neglecting overall factors, is a quadratic equation in $\ln \smin$
\begin{equation} \label{e_xsec_nbody} \end{equation}
\lefteqn{
d\sigma^\txt{\NLOa}_\txt{\nbody} 
	~\propto~ \sum_{ij} \left\{ \hspace{.5cm}
\left[
	1 ~+~ \frac{\alphaS(\mu_R)}{2\pi} \frac{N_C^2-1}{N_C} \left(
			- ( \ln \frac{\smin}{\hat{s}} )^2
			+ \frac{\pi^2}{2} \cdot \frac{1}{3}
			- 2 \cdot \frac{3}{4} \ln \frac{\smin}{\hat{s}}
			- \frac{9}{18} 
	\right)
\right] \right.
} \\ \\
\lefteqn{
\mbox{}\hspace{5cm}	\times ~ |\Matrix^{ij}_\txt{Born} |^2 
f_{P_1 \rightarrow i}(x_1, \mu_F) f_{P_2 \rightarrow j}(x_2, \mu_F)
} \\ \\
\lefteqn{
\mbox{}\hspace{3.25cm}
~+~ \frac{\alphaS(\mu_R)}{2\pi} N_C |\Matrix^{ij}_\txt{Born} |^2 ~
\left[ ~
	f_{P_1 \rightarrow i}(x_1, \mu_F) 
	\left( 
		A_{P_2 \rightarrow j}(x_2, \mu_F)  \ln \frac{\smin}{\mu_F^2} 
		~+~ B^{\MSbar}_{P_2 \rightarrow j}(x_2, \mu_F)
\right) \right.
} \\ \\
\lefteqn{
\mbox{}\hspace{6.5cm}	~+~
	\left. \left. \left( 
		A_{P_1 \rightarrow i}(x_1, \mu_F)  \ln \frac{\smin}{\mu_F^2} 
		~+~ B^{\MSbar}_{P_1 \rightarrow i}(x_1, \mu_F) \right)
	f_{P_2 \rightarrow j}(x_2, \mu_F) 
~ \right] ~ \right\} ,
} \\ \\
where the sum runs over all flavors of initial state (anti)quarks,
$N_C=3$ is the number of quark colors,
$\sqrt{\hat{s}}$ is the vector-boson mass, $\Matrix^{ij}_\txt{Born}$
is the Born level matrix element for 
${\qORqbar}_i {\qORqbar}_j\rightarrow Z^0$, 
$f_{P\rightarrow \qORqbar}(x, \mu_F)$ is the parton density function
evaluated at Bjorken momentum fraction $x$ and factorization scale
$\mu_F$, the renormalization scale is $\mu_R$ (often
$\mu_F=\mu_R=\sqrt{\hat{s}}$ is chosen), and
\begin{equation}
A_{P \rightarrow \qORqbar}(x, \mu_F)  \ln \frac{\smin}{\mu_F^2} 
	~+~ B^{\MSbar}_{P \rightarrow \qORqbar}(x, \mu_F)
\end{equation}
are the crossed structure functions presented in Eq.~3.37 of
Ref.~\cite{Giele:1993dj}.

The solution for $\szero$ corresponds to the smaller of the two roots
of Eq.~\ref{e_xsec_nbody}
\begin{equation}
  \szero ~=~ \mbox{minimum} \left[~
	\hat{s} \times \exp \left( \frac{-b -\sqrt{b^2-4ac}}{2a}
	\right), ~
	\hat{s} \times \exp \left( \frac{-b +\sqrt{b^2-4ac}}{2a}
	\right)~
	\right]
\end{equation}
with
\begin{equation}
\begin{array}{lll}
  a~=~ & \sum_{ij} & - \frac{\alphaS(\mu_R)}{2\pi} \frac{N_C^2-1}{N_C}
	  |\Matrix^{ij}_\txt{Born} |^2 
	  f_{P_1 \rightarrow i}(x_1, \mu_F) f_{P_2 \rightarrow j}(x_2, \mu_F)
\\ \\
  b~=~ &\sum_{ij} &
	-2 \cdot \frac{3}{4} \frac{\alphaS(\mu_R)}{2\pi}
  	\frac{N_C^2-1}{N_C} 
	|\Matrix^{ij}_\txt{Born} |^2 
	f_{P_1 \rightarrow i}(x_1, \mu_F) f_{P_2 \rightarrow j}(x_2,\mu_F) 
\\ \\ && +~
	\frac{\alphaS(\mu_R)}{2\pi} N_C |\Matrix^{ij}_\txt{Born} |^2 ~
	\left[	f_{P_1 \rightarrow i}(x_1, \mu_F) 
		A_{P_2 \rightarrow j}(x_2, \mu_F)
	~+~	A_{P_1 \rightarrow i}(x_1, \mu_F)
		f_{P_2 \rightarrow j}(x_2, \mu_F) \right]
\\ \\
  c~=~&  \sum_{ij} &\hspace{.5cm} \left[
	1 ~+~ \frac{\alphaS(\mu_R)}{2\pi} \frac{N_C^2-1}{N_C} \left(
			+ \frac{\pi^2}{2} \cdot \frac{1}{3}
			- \frac{9}{18} \right) \right]
|\Matrix^{ij}_\txt{Born} |^2 
f_{P_1 \rightarrow i}(x_1, \mu_F) f_{P_2 \rightarrow j}(x_2, \mu_F) 
\\ \\ && +~
	\frac{\alphaS(\mu_R)}{2\pi} N_C |\Matrix^{ij}_\txt{Born} |^2 ~
\left[ 
	f_{P_1 \rightarrow i}(x_1, \mu_F) 
	\left( 
		A_{P_2 \rightarrow j}(x_2, \mu_F)  \ln \frac{\hat{s}}{\mu_F^2} 
		~+~ B^{\MSbar}_{P_2 \rightarrow j}(x_2, \mu_F)
\right) \right.
\\ \\ && \mbox{}\hspace{3cm}  ~+~
	\left. \left( 
		A_{P_1 \rightarrow i}(x_1, \mu_F)  \ln \frac{\hat{s}}{\mu_F^2} 
		~+~ B^{\MSbar}_{P_1 \rightarrow i}(x_1, \mu_F) \right)
	f_{P_2 \rightarrow j}(x_2, \mu_F) 
\right]
\end{array}
\end{equation}


%
%

%
%

%
%

\newcommand{\PSSpicture}{
	\linethickness{3pt}
	\put(10,10){\line(1,0){80}}
	\put(10,10){\line(0,1){80}}
	\put(10,90){\line(1,-1){80}}

	\linethickness{1pt}
	\multiput(16,16)(5,0){14}{\line(1,0){2}}   
	\multiput(16,16)(0,5){14}{\line(0,1){2}}   

	\put(9,80){\makebox(0,0)[tr]{$|Q^2_{1j}|$}}
	\put(80,9){\makebox(0,0)[tr]{$|Q^2_{2j}|$}}
}

\begin{figure}
\setlength{\unitlength}{.086cm}
\noindent
\begin{picture}(100,100)(0,0)
	\PSSpicture
	\linethickness{1pt}
	\put(65,10){\vector(0,1){6}}
	\put(65,16){\vector(0,-1){6}}
	\put(66,13){\makebox(0,0)[l]{$\smin$}}

	\put(25,35){\makebox(0,0)[l]{{\footnotesize\em Resolved Partons}}}
	\put(13,13){\makebox(0,0)[l]{{\footnotesize\em Unresolved Emissions}}}

\end{picture}
\caption{ A projection of the $\ppORppbar\rightarrow Z^0 j$ phase space
onto the $\hat{u}$ vs.\ $\hat{t}$ plane is shown, where
$\hat{u}=(p_2-p_j)^2=-Q^2_{2j}$ and $\hat{t}=(p_1-p_j)^2=-Q^2_{1j}$,
and $p_1,~p_2,~p_j$ are the momenta of the forward colliding parton,
backward colliding parton, and real emission.  The area above
(below) the $\smin$ boundary is the region of resolved
(unresolved) real emissions.  When $\smin=\szero$, it denotes the
boundary defining the region inside of which the the \nbody\ and
\nplusbody\ contributions sum to zero (i.e.\ the cross section integrated
over the unresolved region is zero).  }
\label{f_veto-t_vs_u}
\end{figure}
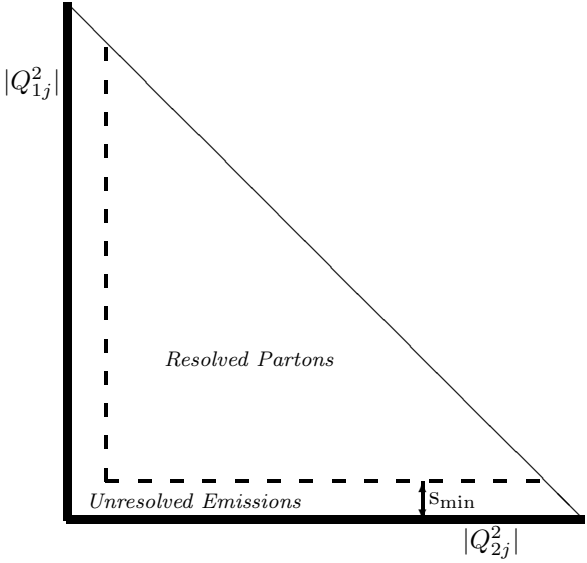

%
%

%
%

%
%

%
%

%
%

%
%

%
%

%
%

%
%

%
%

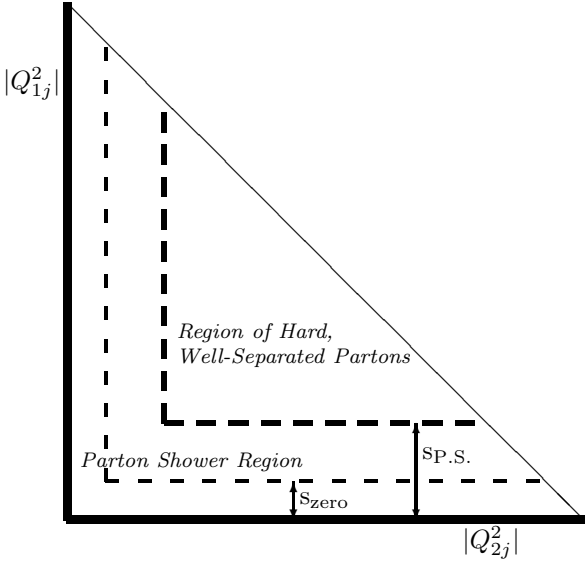
\begin{figure}
\setlength{\unitlength}{.086cm}
\noindent
\begin{picture}(100,100)(0,0)
	\PSSpicture
	\linethickness{1pt}
	\put(45,10){\vector(0,1){6}}
	\put(45,16){\vector(0,-1){6}}
	\put(46,13){\makebox(0,0)[l]{$\szero$}}

	\linethickness{2pt}
	\multiput(25,25)(5,0){10}{\line(1,0){3}}   
	\multiput(25,25)(0,5){10}{\line(0,1){3}}   

	\linethickness{1pt}
	\put(64,10){\vector(0,1){15}}
	\put(64,25){\vector(0,-1){15}}
	\put(65,20){\makebox(0,0)[l]{$\sPS$}}

	\put(12,19){\makebox(0,0)[l]{{\footnotesize\em Parton Shower Region}}}
	\put(27,39){\makebox(0,0)[l]{{\footnotesize\em Region of Hard,}}}
	\put(27,35){\makebox(0,0)[l]{{\footnotesize\em 
						Well-Separated Partons}}}

\end{picture}
\caption{ A projection of the $\ppORppbar\rightarrow Z^0 j$ phase space
onto the $\hat{u}$ vs. $\hat{t}$ plane is shown, where
$\hat{u}=(p_2-p_j)^2=-Q^2_{2j}$ and $\hat{t}=(p_1-p_j)^2=-Q^2_{1j}$,
and $p_1,~p_2,~p_j$ are the momenta of the forward colliding parton,
backward colliding parton, and the hardest emission. Events in the
region of hard well separated partons are sampled with the first order
matrix element, then evolved further by the parton shower. Events in
the region between the $\szero$ and $\sPS$ boundaries are projected
onto \nbody\ kinematics (i.e.\ onto the origin of the plane) and then
evolved with the parton shower to a point which may lie anywhere
below the $\sPS$ boundary. The region below $\szero$ is never
sampled, but may be reached by the projected and showered events.
}
\label{f_vetoPS-t_vs_u}
\end{figure}

%
%

%
%

%
%

%
%

\begin{table}
\caption{ A comparison of computer processing time for the 
\PSveto\ method
and for \PYTHIA. In each case 10000 events are generated for the
process $p\bar{p}\rightarrow Z^0 + X \rightarrow e^+ e^- + X$ at 2~TeV
and the events are evolved through the \PYTHIA\ shower and
hadronization program.  The \PSveto\ event generator requires some
time to initialize the Bases/Spring grids (i.e.\ ``learn the phase
space''), whereas no initialization time is required for \PYTHIA\
processes. The processing time per event and efficiency are similar.
The computer is a 650 MHz Pentium III.  }
\label{t_computer_time}
\begin{tabular}{ l ccc }
  Method & Time for Grid Initialization 
		& Time for 10000 Events & Efficiency \\ \hline
  \PSveto   & 14.0 s & 70.3 s  & 28\% \\
  \PYTHIA   &  ---   & 68.6 s  & 27\% \\
\end{tabular}
\end{table}
%
%


\begin{references}

\bibitem{Sjostrand:1985xi}
T.~Sj\"ostrand,
Phys.\ Lett.\ {\bf B157}, 321 (1985).
\bibitem{Marchesini:1988cf}
G.~Marchesini and B.~R.~Webber,
Nucl.\ Phys.\ {\bf B310}, 461 (1988).

\bibitem{Sjostrand:2001wi} T.~Sj\"ostrand, P.~Eden, C.~Friberg,
L.~L\"onnblad, G.~Miu, S.~Mrenna and E.~Norrbin,
Comput.\ Phys.\ Commun.\  {\bf 135}, 238 (2001) [hep-ph/0010017];
T. Sj\"ostrand, L. L\"onnblad and S. Mrenna, LU~TP~01-21, 
[hep-ph/0108264].

\bibitem{Corcella:2001bw}
G.~Corcella {\it et al.},
JHEP {\bf 0101}, 010 (2001)
[hep-ph/0011363].

\bibitem{Baer:1999sp}
H.~Baer, F.~E.~Paige, S.~D.~Protopopescu and X.~Tata,
hep-ph/0001086.

\bibitem{Collins:2001fm}
J.~Collins,
arXiv:hep-ph/0110113; 
Y.~Chen, J.~C.~Collins and N.~Tkachuk,
JHEP {\bf 0106}, 015 (2001)
[arXiv:hep-ph/0105291];
J.~C.~Collins,
JHEP {\bf 0005}, 004 (2000)
[arXiv:hep-ph/0001040].

\bibitem{Dobbs:2001bx}
M.~Dobbs and M.~Lefebvre,
Phys.\ Rev.\ {\bf D 63}, 053011 (2001)
[hep-ph/0011206].

\bibitem{Dobbs:2001gb}
M.~Dobbs,
Phys.\ Rev.\ D {\bf 64}, 034016 (2001)
[arXiv:hep-ph/0103174].

\bibitem{Potter:2001an}
B.~P\"otter,
Phys.\ Rev.\ D {\bf 63}, 114017 (2001)
[arXiv:hep-ph/0007172].

\bibitem{Altarelli:1979ub}
G.~Altarelli, R.~K.~Ellis and G.~Martinelli,
Nucl.\ Phys.\ B {\bf 157}, 461 (1979);
J.~Kubar-Andre and F.~E.~Paige,
Phys.\ Rev.\ D {\bf 19}, 221 (1979);
K.~Harada, T.~Kaneko and N.~Sakai,
Nucl.\ Phys.\ B {\bf 155}, 169 (1979)
[Erratum-ibid.\ B {\bf 165}, 545 (1979)];
J.~Abad and B.~Humpert,
Phys.\ Lett.\ B {\bf 80}, 286 (1979);
J.~Abad, B.~Humpert and W.~L.~van Neerven,
Phys.\ Lett.\ B {\bf 83}, 371 (1979);
B.~Humpert and W.~L.~Van Neerven,
Phys.\ Lett.\ B {\bf 84}, 327 (1979)
[Erratum-ibid.\ B {\bf 85}, 471 (1979)];
B.~Humpert and W.~L.~Van Neerven,
Phys.\ Lett.\ B {\bf 85}, 293 (1979).


\bibitem{Baur:2001ze}
U.~Baur, O.~Brein, W.~Hollik, C.~Schappacher and D.~Wackeroth,
arXiv:hep-ph/0108274.

\bibitem{Ellis:1981wv} R.~K.~Ellis, D.~A.~Ross and A.~E.~Terrano,
Nucl.\ Phys.\ {\bf B178}, 421 (1981).

\bibitem{Catani:1997vz}
S.~Catani and M.~H.~Seymour,
Nucl.\ Phys.\ B {\bf 485}, 291 (1997)
[Erratum-ibid.\ B {\bf 510}, 503 (1997)]
[arXiv:hep-ph/9605323].


\bibitem{Bergman1989} L.\ J.\ Bergmann, {\it Next-to-leading-log QCD
calculation of symmetric dihadron production}, Ph.D.\ thesis, Florida
State University, 1989;
H.~Baer, J.~Ohnemus and J.~F.~Owens,
Phys.\ Rev.\  {\bf D40}, 2844 (1989).

\bibitem{Harris:2001sx}
B.~W.~Harris and J.~F.~Owens,
hep-ph/0102128.

\bibitem{Fabricius:1981sx}
K.~Fabricius, I.~Schmitt, G.~Kramer and G.~Schierholz,
Z.\ Phys.\ C {\bf 11}, 315 (1981);
G.~Kramer and B.~Lampe,
Fortsch.\ Phys.\  {\bf 37}, 161 (1989).


\bibitem{Giele:1992vf}
W.~T.~Giele and E.~W.~Glover,
Phys.\ Rev.\ D {\bf 46}, 1980 (1992).
\bibitem{Giele:1993dj}
W.~T.~Giele, E.~W.~Glover and D.~A.~Kosower,
Nucl.\ Phys.\ B {\bf 403}, 633 (1993)
[arXiv:hep-ph/9302225].

\bibitem{Baer:1991qf}
H.~Baer and M.~H.~Reno,
Phys.\ Rev.\ D {\bf 44}, 3375 (1991);
H.~Baer and M.~H.~Reno,
Phys.\ Rev.\ D {\bf 45}, 1503 (1992).

\bibitem{Potter:2001ej}
B.~P\"otter and T.~Schorner,
Phys.\ Lett.\ B {\bf 517}, 86 (2001)
[arXiv:hep-ph/0104261].

\bibitem{Glover:1995vz}
E.~W.~Glover and M.~R.~Sutton,
Phys.\ Lett.\ B {\bf 342}, 375 (1995)
[arXiv:hep-ph/9410234].

\bibitem{Aurenche:1981tp}
P.~Aurenche and J.~Lindfors,
Nucl.\ Phys.\ B {\bf 185}, 274 (1981). 
A factor $1/k^2$ is missing from the second term of Eq.~8 in this reference.

\bibitem{Kawabata:1995th}
S.~Kawabata,
Comput.\ Phys.\ Commun.\  {\bf 88}, 309 (1995).
The \CXX\ version is provided by private communication from the author.

\bibitem{Lai:1997mg}
H.~L.~Lai {\it et al.},
Phys.\ Rev.\  {\bf D55}, 1280 (1997)
[hep-ph/9606399].


\bibitem{Davies:1985sp}
C.~T.~Davies, B.~R.~Webber and W.~J.~Stirling,
Nucl.\ Phys.\ B {\bf 256}, 413 (1985).

\bibitem{Boos:2001cv}
E.~Boos {\it et al.},
arXiv:hep-ph/0109068.

\bibitem{Corcella:1998rs}
G.~Corcella and M.~H.~Seymour,
Phys.\ Lett.\ {\bf B442}, 417 (1998)
[hep-ph/9809451].

\bibitem{Catani:2001cc}
S.~Catani, F.~Krauss, R.~Kuhn and B.~R.~Webber,
arXiv:hep-ph/0109231;
B.~R.~Webber,
arXiv:hep-ph/0005035;
F.~Krauss, R.~Kuhn and G.~Soff,
J.\ Phys.\ G {\bf 26}, L11 (2000)
[arXiv:hep-ph/9904274].

\bibitem{Andre:1997vh}
J.~Andre and T.~Sj\"ostrand,
Phys.\ Rev.\ D {\bf 57}, 5767 (1998)
[arXiv:hep-ph/9708390].

\bibitem{Krauss:1999fc}
F.~Krauss, R.~Kuhn and G.~Soff,
Acta Phys.\ Polon.\ B {\bf 30}, 3875 (1999)
[arXiv:hep-ph/9909572];
R.~Kuhn, F.~Krauss, B.~Ivanyi and G.~Soff,
Comput.\ Phys.\ Commun.\  {\bf 134}, 223 (2001)
[arXiv:hep-ph/0004270].


%
\bibitem{Miu:1999ju}
G.~Miu and T.~Sj\"{o}strand,
Phys.\ Lett.\  {\bf B449}, 313 (1999)
[hep-ph/9812455];
T.~Sj\"ostrand,
hep-ph/0001032.
In this reference the \PYTHIA\ prediction is compared to experimental
data. The authors realized after publication that the experimental
data had not been unfolded from detector effects, and so any
interpretations of the comparison should be made with this in mind.

\end{references}
\end{document}